\documentclass{aastex63}
\usepackage{amsmath}
\usepackage{xcolor}
\usepackage[scr=rsfso,frak=euler,bb=ams]{mathalfa}
\newcommand{\figpath}{.}

\begin{document}

\title{Multi-Frequency General Relativistic Radiation-Hydrodynamics with $\bf{M}_1$ Closure}
\shortauthors{Anninos, et al.}
\author{Peter Anninos}
\affil{Lawrence Livermore National Laboratory, P.O. Box 808, Livermore, CA 94550, USA}
\author{P. Chris Fragile}
\affil{Department of Physics \& Astronomy, College of Charleston, Charleston, SC 29424, USA}
\email{fragilep@cofc.edu}

\begin{abstract}
We report on recent upgrades to our general relativistic radiation-magnetohydrodynamics code,
{\em Cosmos++}, which expands the two-moment, $\bf{M}_1$, radiation treatment from grey to multi-frequency transport,
including Doppler and gravitational frequency shifts. The solver accommodates either photon (Bose-Einstein)
or neutrino (Fermi-Dirac) statistical distribution functions with absorption, emission,
and elastic scattering processes. An implicit scheme is implemented to
simultaneously solve the primitive inversion problem together with the
radiation-matter coupling source terms, providing stability over a broad range of opacities
and optical depths where the interactions terms can be stiff. 
We discuss our formulations and numerical methods, and validate our methods against
a wide variety of test problems spanning optically thin to thick regimes in flat, weakly curved,
and strongly curved spacetimes.
\end{abstract}
\keywords{Computational methods --- Radiative magnetohydrodynamics}

\section{Introduction}
\label{sec:intro}

Today we are part of an exciting multi-messenger era in astronomy. Telescopes cover the entire electromagnetic (EM) spectrum, with nightly full-sky coverage becoming a reality \citep[e.g.,][]{panstarrs16,lsst19,ztf19}. Additionally, cosmic ray detectors \citep[e.g.,][]{cherenkov19}, neutrino detectors \citep[e.g.,][]{icecube19}, and now gravitational wave detectors \citep{abbott18} give us views of the universe beyond EM radiation, and as each new means of observation has been added, new discoveries have quickly followed \citep[e.g.,][]{abbott17,metzger17,icecube18,keivani18,fang19}. Certainly many more are to be expected.

A principle focus of multi-messenger astronomy is the transient universe \citep{charles13}, particularly events that are characterized by short bursts of electromagnetic radiation, possibly accompanied by cosmic ray, neutrino, or gravitational wave signals, such as kilonovae \citep[e.g.,][]{abbott17,metzger17}, fast radio bursts \citep[e.g.,][]{burke18,wang20}, gamma ray bursts \citep[e.g.,][]{burns19}, and tidal disruption events \citep[e.g.,][]{senno17}. These events are often highly energetic and commonly associated with compact objects (white dwarfs, neutron stars, or black holes), suggesting relativistic physics plays a role.

The many new discoveries in multi-messenger astronomy need to be matched by corresponding developments in the computational tools that help in their interpretation and understanding. Over the decades, advances in observational capabilities have seen parallel developments in astrophysical simulation tools toward ever higher levels of sophistication, starting from relatively simple hydrodynamic and N-body simulations to magnetohydrodynamics (MHD), radiation MHD, and beyond \citep[see][for a review of relativistic code development]{abramowicz13}. Most transient phenomena require some combination of relativity, hydrodynamics, magnetic fields, and radiation to be adequately modeled. Fortunately, the number of codes available for advanced radiation MHD simulations has quite literally exploded in recent years \citep[an incomplete list includes][]{farris08,muller10,shibata11,zanotti11,jiang12,lentz12,sadowski13,zhang13,gonzalez15,just15,tominaga15,kuroda16,skinner19,ryan20,weih20}. Our own contribution is the general relativistic radiation magnetohydrodynamics code, {\sc Cosmos++} \citep{anninos05,fragile12,fragile14}, which includes a discontinuous-Galerkin variant, {\sc CosmosDG} \citep{anninos17}.

As ever, though, numerical simulations are only an approximation to reality. The current limitation in radiation MHD is that solving the full Boltzmann transport equation remains computationally challenging and not entirely practical in most scenarios, owing to the large number of degrees of freedom (in space and frequency) and wide range of optical depths, although new formulations have been developed for this purpose \citep{davis20}. Therefore, most codes today treat some simplified form of radiation. One common approach is to use a scheme where only the first few moments of the radiation distribution function are evolved \citep{thorne81,shibata11}. The most basic is the flux-limited diffusion approximation \citep{levermore81,pomraning81}, which only treats the zeroth moment, meaning it retains information only about the radiation intensity, but not the direction of its propagation. A two-moment scheme, such as $\bf{M}_1$ \citep{levermore84,dubroca99,gonzalez07}, retains both the intensity and (average) direction of radiation flow, yet still closes the system of equations at a level that remains computationally reasonable. This approach has seen wide implementation in the context of black-hole accretion \citep{sadowski13,fragile14,mckinney14,mishra16,takahashi16,fragile18}, black-hole--neutron-star mergers \citep{foucart15,foucart16b}, binary neutron stars \citep{foucart16a,sekiguchi16}, core-collapse supernovae \citep{oconnor15,just15,kuroda16}, and the interaction of Type I X-ray bursts with accretion disks \citep{fragile18b,fragile20}.

Another simplification is that most relativistic radiation MHD codes today assume a frequency-integrated (or grey) opacity and evolve the radiation field with a single characteristic frequency. However, resolving the photon (or neutrino) frequency (energy), even crudely, can be crucially important to properly modeling and understanding many transient phenomena, such as core-collapse supernovae \citep[e.g.,][]{janka12,burrows13,foglizzo15}, tidal disruption events \citep{dai18}, and the outbursts of black hole X-ray binaries.

This is the goal of our current work, to extend our radiation transport capabilities 
from a frequency-integrated (grey) approximation to a multi-frequency (equivalently multi-energy or sometimes called multi-group) method by discretizing
the radiation energy and flux equations in frequency as well as space and time. As in our previous
paper \citep{fragile14}, we adopt the $\bf{M}_1$ closure for general relativistic transport,
though we have additionally developed a multi-frequency, flux-limited diffusion (with an anisotropic Eddington tensor) solver for Newtonian systems.
Much of the methodology, including the formalism and numerical methods, discussed in this paper is taken from
\citet{fragile14}, and we occasionally refer the reader back to that paper for further details, particularly regarding
the high resolution shock capturing algorithms and the primitive inversion scheme. All of those specifics
are similar to what we have developed for this work, except they are applied to the radiation fields in each
frequency bin separately. The primitive inversion scheme utilizes a first-order Taylor expansion of the
conserved fields together with the radiation coupling terms, again similar to our previous work except
here the dimension of the matrix system scales with the number of radiation bins, and the coupling
to the hydrodynamics occurs after integrating each source contribution over frequency.
The most significant new element that comes from frequency-dependent transport is the introduction
of a source term responsible for advecting energies in frequency space as radiation propagates 
through gravitational fields or experiences fluid velocities that produce shifts in the photon frequencies (or neutrino energies).

As for the organization of this paper, Section \ref{sec:formalism} follows with
an overview of the essential formalism and conservation equations.
Section \ref{sec:methods} discusses our numerical implementation with an emphasis on the new elements: frequency advection,
closure relations, and the implicit approach for solving the coupled primitive inversion and 
multi-frequency radiation source terms.  
Section \ref{sec:tests} reports on a series of validating test problems, and 
we conclude in Section \ref{sec:conclusion}.

Most of the equations in this paper are written in units where $G = c = 1$, 
although in a few places we leave in factors of $c$ for clarity. 
We adopt the usual convention whereby Greek (Latin) indices refer to spacetime (spatial)
coordinates and adopt a $(-,+,+,+)$ metric signature.

\section{Formalism}
\label{sec:formalism}

A multi-frequency treatment of radiation transport can be derived by selecting a finite set of frequency groups (or bins)
and defining the discrete energy densities $E_{(\nu)n}$ as integrals of the energy spectral densities, $E_{(\nu)}$,
over the group frequency interval $\delta\nu_n$. Mathematically, $E_n = \int_{\nu_n-\delta\nu_n/2}^{\nu_n+\delta\nu_n/2} E_{(\nu)} d\nu$, and
therefore $E=\sum_n E_{(\nu)n} = \int_{\nu_l}^{\nu_u} E_{(\nu)} d\nu$, where $\nu_l$ and $\nu_u$ are the lower and upper limits to our frequency bins. In this notation, the total stress-energy
tensor can be written
\begin{equation}
T^{\alpha\beta} = T^{\alpha\beta}_{\text{fluid}} + \int d\nu \sum_i R^{\alpha\beta}_{i\ (\nu)} ~,
\end{equation}
where $T^{\alpha\beta}_{\text{fluid}}$ is the fluid component, and $ R^{\alpha\beta}_{i\ (\nu)}$ is the spectral
radiation stress tensor summed over all radiation components $i$ representing photons or different neutrino species.
For this work we consider only single species (either photon or single flavor neutrino) transport.

The spectral radiation tensor, $R^{\alpha \beta}_{(\nu)}$, can be written in any number of ways, depending on the frame of reference.
For example, the following representations
\begin{eqnarray}
R^{\alpha \beta}_{(\nu)} &=& E_{(\nu)} n^\alpha n^\beta + F^\alpha_{(\nu)} n^\beta + F^\beta_{(\nu)} n^\alpha + P^{\alpha \beta}_{(\nu)} ~, \\
                         &=& J_{(\nu)} u^\alpha u^\beta + H^\alpha_{(\nu)} u^\beta + H^\beta_{(\nu)} u^\alpha + L^{\alpha \beta}_{(\nu)} ~, \\
                         &=& \frac{4}{3} E_{R(\nu)} u^{\alpha}_{R(\nu)} u^{\beta}_{R(\nu)} + \frac{1}{3} E_{R(\nu)} g^{\alpha \beta} ~,
\label{eqn:radstress_all}
\end{eqnarray}
are commonly used for Eulerian (lab), co-moving (fluid), or isotropic (radiation) frame formalisms, respectively, where 
$n_\alpha = (-\alpha,\ 0,\ 0,\ 0)$ is a timelike vector orthogonal to the spacelike hypersurface, $u^\alpha$ is the fluid rest frame 4-velocity, and $\alpha = 1/\sqrt{-g^{00}}$ is the lapse function. The quantities $E_{(\nu)}$, $J_{(\nu)}$, and $E_{R(\nu)}$ represent the frequency-dependent radiation energy densities in the different frames; likewise, $F^\alpha_{(\nu)}$ and $H^\alpha_{(\nu)}$ represent the radiation momentum densities. Finally, $P^{\alpha \beta}_{(\nu)}$ and $L^{\alpha \beta}_{(\nu)}$ are most often referred to as the pressure or stress tensors of the radiation.
In previous work \citep{fragile14}, we adopted the radiation frame formalism \citep{sadowski13},
which offers some unique advantages for computation.
Notice, for example, that the radiation pressure does not appear in the isotropic stress-tensor,
as it implicitly represents the covariant formulation of the  $\bf{M}_1$ closure scheme \citep{levermore84},
which assumes the radiation is isotropic in the radiation rest frame\footnote{We are as yet unaware of any general proof of the existence of such a frame, though from our experience we have not found any cases where this formulation breaks down.}. There is also no explicit appearance of a radiation momentum density in the radiation frame.
However, certain calculations are most conveniently done in either of the fluid or Eulerian
frames, so we retain the flexibility to transform between the different reference frames and fields. To this
end, we explicitly write out here some of the more important relations between radiation fields and moments.

First, the frame-dependent primitive moments are extracted from the spectral 
radiation stress-tensor (or equivalently the conserved or evolved radiation fields) as
\begin{equation}
E_{(\nu)}      =  R^{\alpha\beta}_{(\nu)} n_\alpha n_\beta ~, \quad
F^i_{(\nu)}    = -R^{\alpha\beta}_{(\nu)} n_\alpha \gamma^i_\beta ~, \quad
P^{ij}_{(\nu)} =  R^{\alpha\beta}_{(\nu)} \gamma^i_\alpha \gamma^j_\beta ~, 
\label{eqn:moments}
\end{equation}
and 
\begin{equation}
J_{(\nu)}          =  R^{\alpha\beta}_{(\nu)} u_\alpha u_\beta ~, \quad
H^\gamma_{(\nu)}   = -R^{\alpha\beta}_{(\nu)} u_\alpha h^\gamma_\beta ~, \quad
L^{\gamma\delta}_{(\nu)} =  R^{\alpha\beta}_{(\nu)} h^\gamma_\alpha h^\delta_\beta ~, 
\end{equation}
where $\gamma_{\alpha\beta} = g_{\alpha\beta} + n_\alpha n_\beta$ is the spatial metric,
$h_{\alpha\beta} = g_{\alpha\beta} + u_\alpha u_\beta$ is the fluid-frame projection metric, and the second rank tensors $P^{ij}_{(\nu)}$ and $L^{\gamma\delta}_{(\nu)}$ are determined by closure relations 
to be discussed later. Additionally we can relate the energy and flux variables directly via
\citep{shibata11}
\begin{eqnarray}
J_{(\nu)} &=& E_{(\nu)} w^2 - 2 w F^k_{(\nu)} u_k + P^{ij}_{(\nu)} u_i u_j ~, \label{eq:Jnu}\\
H^\alpha_{(\nu)} &=& \left[E_{(\nu)} w - F^k_{(\nu)} u_k\right] h^\alpha_\beta n^\beta
                 + w h^\alpha_\beta F^\beta_{(\nu)} - P^{ij}_{(\nu)} h^\alpha_i u_j ~, \label{eq:Halphanu}
\end{eqnarray}
where
$w=\alpha u^0$ is the Lorentz factor.
The Eulerian frame flux vector and pressure tensor additionally satisfy $F^\alpha n_\alpha = P^{\alpha\beta} n_\alpha= 0$,
implying $F^0_{(\nu)} = P^{0\alpha}_{(\nu)} = 0$, a fact we have exploited in writing Equations (\ref{eq:Jnu}) and (\ref{eq:Halphanu}).

The radiation variables, $E_{R(\nu)}$ and $u^\alpha_{R(\nu)}$, representing  the 
spectral radiation energy density in the radiation rest frame and the 4-velocity of the radiation rest frame itself,
can easily be defined in terms of either
lab or fluid frame tensor components. In particular, the following quadratic equations
\begin{eqnarray}
g_{\alpha\beta} R^{0 \alpha}_{(\nu)} R^{0 \beta}_{(\nu)} & = & -\frac{8}{9} E_{R(\nu)}^2 \left[u^0_{R(\nu)}\right]^2 + \frac{1}{9} E_{R(\nu)}^2 g^{00} ~, \label{eqn:rtt1} \\
R^{00}_{(\nu)} & = & \frac{4}{3} E_{R(\nu)} \left[u^0_{R(\nu)}\right]^2 + \frac{1}{3} E_{R(\nu)} g^{00} ~,
\label{eqn:rtt2}
\end{eqnarray}
can be solved for $ E_{R(\nu)}$ and $u^0_{R(\nu)}$ \citep{sadowski13}. The remaining spatial components of the
radiation 4-velocity, $u^i_{R(\nu)}$, are derived from the time components of the radiation stress-tensor.

Following the truncated moment formalism \citep{thorne81, shibata11}, the radiation conservation equations become
\begin{equation}
R^{\alpha\beta}_{(\nu) ; \beta} 
   - \frac{\partial}{\partial\nu}\left[\nu M^{\alpha\beta\gamma}_{(\nu)} u_{\beta ;\gamma}\right]
   = -G_{\alpha (\nu)}   ~,
\end{equation}
where $ G_{\alpha (\nu)}$ represents radiation-matter interaction source terms, and $M^{\alpha\beta\gamma}_{(\nu)}$
is the third-rank moment tensor associated with Doppler and gravitational frequency shifts.

These radiation equations are solved together with the conservation equations for mass
$(\rho u^\beta)_{;\beta} = 0$ and fluid stress-energy $(T^\beta_{\ \alpha})_{;\beta} = \int G_{\alpha (\nu)} d\nu$.
Ignoring non-ideal effects and magnetic fields, the fluid stress-energy tensor takes the form
\begin{equation}
T^{\alpha \beta} = (\rho + \rho \epsilon + P_\mathrm{gas}) u^\alpha u^\beta + P_\mathrm{gas} g^{\alpha \beta} ~,
\end{equation}
where $P_\mathrm{gas}$ is the gas pressure.  
Although we do not consider magnetic fields or viscosity in this work, we advertise that both of these physics capabilities
are currently fully integrated with this multi-frequency radiation upgrade. We refer the reader to 
\citet{fragile12} and \citet{fragile18} for details on their respective implementations.

Coupling of the fluid and radiation equations occurs through the radiation 4-force density, $G^\mu_{(\nu)}$, 
written in the form 
\begin{equation}
G^\mu_{(\nu)} = -\rho \left[\kappa^{\mathrm{a}}_{(\nu)} + \kappa^{\mathrm{s}}_{(\nu)}\right] R^{\mu \nu }_{(\nu)} u_{\nu} 
                -\rho\left[\kappa^{\mathrm{s}}_{(\nu)} R^{\alpha \beta}_{(\nu)} u_{\alpha} u_{\beta} + 
                           \kappa^{\mathrm{a}}_{(\nu)} B_{(\nu)} \right] u^{\mu}~,
\label{eq:Galpha}
\end{equation}
where $\kappa^\mathrm{a}_{(\nu)}$ and $\kappa^\mathrm{s}_{(\nu)}$ represent the 
frequency-dependent absorption/emission and elastic scattering opacities, 
respectively; $B_{(\nu)}$ is the Bose-Einstein or Fermi-Dirac statistical distribution function
\begin{equation}
B_{(\nu)} = \frac{4\pi g (h\nu)^3}{(hc)^3}\left(\frac{1}{e^{(h\nu-\mu_\nu)/(kT)} - \eta}\right)  ~,
\end{equation}
where $\mu_\nu$ is the chemical potential,
$g = 2$ (1) is the statistical weight for photons (neutrinos), and $\eta = 1$ (-1) for photons (neutrinos).
The grey (frequency integrated) version of equation (\ref{eq:Galpha}) can be written
\begin{equation}
G^\mu= -\rho \left(\kappa^{\mathrm{a}}_{\mathrm{F}} + \kappa^{\mathrm{s}} \right) R^{\mu \nu } u_{\nu} 
       -\rho\left[(\kappa^{\mathrm{s}} + \kappa^{\mathrm{a}}_{\mathrm{F}} - 
                   \kappa^{\mathrm{a}}_{\mathrm{A}}) R^{\alpha \beta} u_{\alpha} u_{\beta} + 
                   \kappa^{\mathrm{a}}_{\mathrm{P}} a_R T^4 \right] u^{\mu}~,
\label{eq:GalphaSG}
\end{equation}
where $a_R$ is the radiation constant (different for photons and neutrinos) and $\kappa^{\mathrm{a}}_{\mathrm{F}}$,
$\kappa^{\mathrm{a}}_{\mathrm{A}}$, and $\kappa^{\mathrm{a}}_{\mathrm{P}}$ are the flux, absorption, and Planck
mean opacities, respectively.

Expanding out the covariant derivatives, the full set of conservation equations to be solved are written as
\begin{equation}
 \partial_t D + \partial_i (DV^i)  =  0 ~,  \label{eqn:de}
\end{equation}
\begin{equation}
\partial_t {\cal E} + \partial_i \left(-\sqrt{-g}~T^i_0\right)  = 
      -\sqrt{-g}~T^\alpha_\beta~\Gamma^\beta_{0 \alpha} - \sqrt{-g}~G_{0} ~,
    \label{eqn:en2} 
\end{equation}
\begin{equation}
 \partial_t {\cal S}_j + \partial_i \left(\sqrt{-g}~T^i_j\right)  = 
      \sqrt{-g}~T^\alpha_\beta~\Gamma^\beta_{j \alpha} + \sqrt{-g}~G_{j} ~,
    \label{eqn:mom2} 
\end{equation}
\begin{equation}
 \partial_t {\cal R}_{(\nu)} + \partial_i \left[-\sqrt{-g}~R^i_{0 (\nu)}\right]  = 
     - \frac{\partial}{\partial\nu}\left[\nu M_{0\beta\gamma (\nu)} u^{\beta ;\gamma}\right]
     - \sqrt{-g}~R^\alpha_{\beta (\nu)}~\Gamma^\beta_{0 \alpha} + \sqrt{-g}~G_{0 (\nu)} ~,
    \label{eqn:rad_en} 
\end{equation}
\begin{equation}
 \partial_t {\cal R}_{j (\nu)} + \partial_i \left[\sqrt{-g}~R^i_{j (\nu)}\right]  = 
     \frac{\partial}{\partial\nu}\left[\nu M_{j\beta\gamma (\nu)} u^{\beta ;\gamma}\right]
     + \sqrt{-g}~R^\alpha_{\beta (\nu)}~\Gamma^\beta_{j \alpha} - \sqrt{-g}~G_{j (\nu)} ~,
    \label{eqn:rad_mom} 
\end{equation}
where $D=W\rho$, $\rho$ is the rest-frame fluid density, $W=\sqrt{-g} u^0 = \sqrt{-g}(\gamma/\alpha)$ is 
the relativistic boost factor, $V^i=u^i/u^0$ is the fluid transport velocity, 
$g$ is the 4-metric determinant, $\Gamma^\beta_{\alpha\gamma}$ is the geometric connection coefficients of the metric,
${\cal E} = -\sqrt{-g} T^0_0$ is the total energy density, ${\cal S}_j = \sqrt{-g} T^0_j$ is the 
covariant momentum density, ${\cal R}_{(\nu)} = -\sqrt{-g} R^0_{0 (\nu)}$ is the conserved radiation spectral energy,
${\cal R}_{j (\nu)} = \sqrt{-g} R^0_{j (\nu)}$ is the conserved radiation spectral momentum, and
$G_\alpha= \int G_{\alpha (\nu)} d\nu$.

\section{Numerical Implementation}
\label{sec:methods}

Equations (\ref{eqn:de}) - (\ref{eqn:rad_mom}) are solved by operator splitting terms into
spacetime advection, curvature, frequency advection, and radiation-matter coupling.
The first three contributions are solved using high-order explicit methods, while the fourth is updated with a fully implicit approach,
which provides stability when radiation-matter interactions become stiff relative to a hydrodynamic
time scale, as they often do when strongly coupled. Solution methods for each of these contributions are discussed below.

\subsection{Advection and Curvature}
\label{subsec:advection}

The radiation conservation laws (\ref{eqn:rad_en}) - (\ref{eqn:rad_mom}) 
are identical in form to the fluid energy and momentum conservation equations
already solved in {\em Cosmos++}, and are amenable to similar numerical techniques, specifically the 
high-resolution shock-capturing (HRSC) scheme as described in \citet{fragile12}.  

Representing conserved fields as 
$\mathbf{U} = [D, \ {\cal E}, \ {\cal S}_j, \ {\cal R}_{(\nu)}, \ {\cal R}_{j(\nu)}]$,
the discrete finite volume representation of equations (\ref{eqn:de}) - (\ref{eqn:rad_mom}) 
are written in generic fashion as
\begin{equation}
\mathbf{U}^* = \mathbf{U}^n - \frac{\Delta t}{V} \sum\limits_{\text{cell\ faces}}\left(\mathbf{F}^i A_i\right)^n + \Delta t ~\mathbf{S}_c^n ~,
\end{equation}
where $\mathbf{S}_c(\mathbf{P})$ contains the curvature source terms,
$\mathbf{F}^i(\mathbf{P}) = \sqrt{-g} [\rho u^0,\ -T^i_0,\ T^i_j,\ -R^i_{0(\nu)},\ R^i_{j(\nu)}]$ 
are the fluxes, and $\mathbf{U}^*$ represents the intermediate solution state
(accounting for advection and curvature, but not frequency shift or coupling terms).
Notice that both the flux and curvature source terms are computed from the set of primitive, not conserved, fields.

One of the differences between this work and that presented in \cite{fragile14} is the choice of primitive fields.
Here we have opted to use the fluid and spectral radiation 4-velocities [$u^i$ and $u^i_{R(\nu)}$] 
and not the normal observer projected 4-velocities
($\tilde u^i = u^i - u^0 g^{0i}/g^{00}$) that we used previously.
With this change, the set of primitive variables becomes
$\mathbf{P} = [\rho,\ \epsilon,\ {u}^i,\ E_{R(\nu)},\ {u}^i_{R(\nu)}]$,
where $\epsilon$ is the specific internal energy as measured in the fluid rest frame. 

The flux terms are calculated at zone faces using either the Harten-Lax-van Leer (HLL) or 
Lax-Friedrichs (LF) Riemann solver with options for linear or PPM slope limited reconstruction of the primitive fields.
For the HLL solver this takes the form
\begin{equation}
\mathbf{F}_\mathrm{HLL} = \frac{\lambda_+ \mathbf{F}_\mathrm{L} - \lambda_- \mathbf{F}_\mathrm{R} + \lambda_- \lambda_+(\mathbf{U}_\mathrm{R} - \mathbf{U}_\mathrm{L})}
                        {\lambda_+ - \lambda_-}  ~,
\end{equation}
where R (L) subscripts denote right (left) reconstructed states, and 
$\lambda_+$ ($\lambda_-$) is the characteristic maximum (minimum) wave speed.

One of the advantages of formulating radiation transport in 
terms of the primitive radiation variables, $E_R$ and $u^i_R$, 
is that it simplifies the calculation of characteristic radiation
wave speeds required for the Riemann solvers.  We generally
follow the prescription outlined in our grey treatment \citep{fragile12} where we effectively
replace the fluid velocity with the radiation velocity in the co-moving dispersion
relation \citep{gammie03} \citep[see also][]{mckinney14}
\begin{equation}
\left[1-v_T^2\left(1 + \frac{g^{00}}{u^0 u^0}\right)\right] (c^i)^2 +
2\left[v_T^2(\left(V^i + \frac{g^{0i}}{u^0 u^0}\right) - V^i\right] c^i 
+ \left[V^i V^i - v_T^2\left(V^i V^i + \frac{g^{ii}}{u^0 u^0}\right)\right] = 0 ~,
\label{eqn:wavespeed}
\end{equation}
where $c^i$ is the wave speed along each coordinate direction, $x^i$, and
$v_T$ is the maximum of the fluid or radiation wave speeds ($1/\sqrt{3}$ in the optically thin regime).
The minimum ($\lambda_-$) and maximum ($\lambda_+$) speeds are defined by the minimum and maximum
solutions of the quadratic equation (\ref{eqn:wavespeed}).
Generalization to the optically thick regime is accommodated by limiting the characteristic
velocities as
\begin{eqnarray}
\lambda_- &\rightarrow& \text{max} \left(\lambda_-, ~ -\frac{4}{3\tau}\right) \\
\lambda_+ &\rightarrow& \text{min} \left(\lambda_+, ~  \frac{4}{3\tau}\right) 	~,
\end{eqnarray}
where $\tau$ is the total optical depth in the cell.
Although it appears to make little difference, we provide an
alternative extension of the wavespeed into the optically thick regime by
using the fluid-frame moments and the limiting procedure described
in section \ref{subsec:doppler} to interpolate between the two
\begin{equation}
\lambda_\pm = \frac{3\chi-1}{2} \lambda_{\pm,\mathrm{thin}} + \frac{3(1-\chi)}{2} \lambda_{\pm,\mathrm{thick}}  ~,
\end{equation}
where $\lambda_{\pm,\mathrm{thin}}$ and $\lambda_{\pm,\mathrm{thick}}$ are the corresponding speeds in the optically thin
and thick regimes respectively. 

Advection and curvature operators are completed (advanced) with several available time discretization options. 
{\em Cosmos++} supports numerous options designed to enhance stability and accuracy for specific applications
and algorithms, including (up to) fourth order
strong-stability preserving Runge-Kutta methods that benefit high order finite elements
\citep{anninos17}, and multi-step Crank-Nicholson methods that stabilize highly dynamical
black hole spacetimes. For the test problems presented in this report we typically use a more conventional
second order time discretization based on a low-storage forward Euler method.

\subsection{Doppler and Gravitational Frequency Shifts}
\label{subsec:doppler}

The frequency advection source terms are updated following the general procedure outlined in 
\citet{shibata11} \citep[see also][]{kuroda16}, after transforming the radiation-frame moments to their lab frame
counterparts using equation (\ref{eqn:moments}). The conservation equations for the lab frame moments take the form
\begin{eqnarray}
 \partial_t {E}_{(\nu)}   &=&
     - \frac{\partial}{\partial\nu}\left[\nu n_\alpha M^{\alpha\beta\gamma}_{(\nu)} u_{\beta ;\gamma}\right] ~,
    \label{eqn:dop_en}  \\
 \partial_t {F}_{j (\nu)} &=&
       \frac{\partial}{\partial\nu}\left[\nu \gamma_{j\alpha} M^{\alpha\beta\gamma}_{(\nu)} u_{\beta ;\gamma}\right] ~,
    \label{eqn:dop_mom} 
\end{eqnarray}
where 
\begin{equation}
M^{\alpha\beta\gamma}_{(\nu)} u_{\gamma ;\beta} =   \left[
                                    H^\gamma_{(\nu)} u^\alpha u^\beta
                                  + L^{\alpha\gamma}_{(\nu)} u^\beta
                                  + L^{\beta\gamma}_{(\nu)} u^\alpha
                                  + N^{\alpha\beta\gamma}_{(\nu)} 
                              \right]  u_{\gamma ;\beta} ~,
\end{equation}
and $N^{\alpha\beta\gamma}_{(\nu)}$ is determined by a closure formulation connecting
optically thin and thick regimes
\begin{equation}
N^{\alpha\beta\gamma}_{(\nu)} = \frac{3\chi-1}{2}   \left[N^{\alpha\beta\gamma}_{(\nu)}\right]_{\text{thin}}
                              + \frac{3(1-\chi)}{2} \left[N^{\alpha\beta\gamma}_{(\nu)}\right]_{\text{thick}} ~.
\end{equation}
We adopt the thin/thick expressions recommended in \citet{shibata11}
\begin{equation}
\left[N^{\alpha\beta\gamma}_{(\nu)}\right]_{\text{thin}} =
   \frac{J_{(\nu)} H^\alpha_{(\nu)} H^\beta_{(\nu)} H^\gamma_{(\nu)}}
        {(h_{\alpha\beta} H^\alpha_{(\nu)} H^\beta_{(\nu)})^{3/2}}
\end{equation}
and
\begin{equation}
\left[N^{\alpha\beta\gamma}_{(\nu)}\right]_{\text{thick}} = 
    \frac{1}{5}\left[ H^\alpha_{(\nu)} h^{\beta\gamma} +
                      H^\beta_{(\nu)}  h^{\alpha\gamma} +
                      H^\gamma_{(\nu)} h^{\alpha\beta} \right] ~.
\end{equation}
Among the many options for the closure function, $\chi$, we have chosen to use \citep{levermore84}
\begin{equation}
\chi = \frac{3 + 4\xi^2}{5 + 2 \sqrt{4 - 3\xi^2}} ~,
\label{eqn:chi}
\end{equation}
where
\begin{equation}
\xi^2 = \frac{h_{\alpha\beta} H^\alpha_{(\nu)} H^\beta_{(\nu)}}
                     {J^2_{(\nu)}}
\label{eqn:xi}
\end{equation}
works well as an indicator of whether the fluid is locally optically thick ($\xi \rightarrow 0$)
or thin ($\xi \rightarrow 1$).

The form of equations (\ref{eqn:dop_en}) and (\ref{eqn:dop_mom}) are advective
in nature and fully conservative
when the boundary conditions enforce zero radiation flux at the edges of the frequency
domain. We thus discretize and update
both equations using a conservative multi-stage, second order upwind scheme where the flux
terms are reconstructed at group boundaries using a minmod limiter to preserve
monotonicity in the gradient extrapolants. The scheme is multi-stage in the sense
that we subcycle the source update, respecting the characteristic
advection time for the most rapidly changing bin energies. In particular, the
subcycle timestep is determined by the minimum advection time over all groups
based on the covariant divergence of the fluid 4-velocity,
$\delta t_\mathrm{sub} = \text{min}[C_\mathrm{cfl} \delta\nu/(\nu |u^\alpha_{; \alpha}|)]$,
where $C_\mathrm{cfl} < 1$ is a Courant factor typically set to 0.3.
After advancing the lab-frame moments with equations (\ref{eqn:dop_en}) and (\ref{eqn:dop_mom}), 
the evolved radiation fields are easily reconstructed from the radiation stress-energy tensor
(\ref{eqn:radstress_all}) along with the following general relativistic $\bf{M}_1$ closure relation for the pressure:
\begin{equation}
P^{ij}_{(\nu)} = \frac{3\chi-1}{2}   \left[P^{ij}_{(\nu)}\right]_{\text{thin}}
                              + \frac{3(1-\chi)}{2} \left[P^{ij}_{(\nu)}\right]_{\text{thick}} ~,
\end{equation}
where
\begin{equation}
\left[P_{(\nu)}\right]^{ij}_{\text{thin}} = E_{(\nu)} 
                           \frac{F_{(\nu)}^i F_{(\nu)}^j}
                                {\gamma_{ij} F_{(\nu)}^i F_{(\nu)}^j} ~,
\end{equation}
and $\left[P_{(\nu)}\right]^{ij}_{\text{thick}} = E_{(\nu)} \gamma^{ij}/3$.

\subsection{Primitive Inversion and Radiation-Matter Coupling}
\label{sec:primitive}

The radiation-matter coupling terms are updated using an implicit, iterative Newton-Raphson method
to provide greater stability in strongly coupled regimes. In this section, we will use the index $n$ to indicate steps or cycles in the global time-stepping scheme, whereas the index $m$ indicates iteration steps within the Newton-Raphson solver. Once all explicit steps
have been advanced, the implicit solve follows according to
\begin{equation}
\mathbf{U}^{n+1} = \mathbf{U}^* + \Delta t ~\mathbf{S}_r^{n+1} ~,
\label{eqn:iterate}
\end{equation}
where 
$\mathbf{S}_r(\mathbf{P})^{n+1} = \sqrt{-g} \left[ 0,\ -G_0,\ G_j,\ G_{0(\nu)},\ -G_{j(\nu)} \right]^{n+1}$
represents the interactions terms at the advanced time $n+1$.
Taking the 1st order Taylor expansion with respect to primitive variables, the ($m+1$)st iterate is approximated as
\begin{eqnarray}
\mathbf{U}^{m+1} &=& \mathbf{U}^m + \sum_a \left(\frac{\partial\mathbf{U}}{\partial P^a}\right)^m \delta P^a \\
G_\alpha^{m+1} &=& G_\alpha^m + \sum_a \left(\frac{\partial G_\alpha}{\partial P^a}\right)^m \delta P^a ~,
\end{eqnarray}
where
\begin{eqnarray} 
\delta \mathbf{P} = \left(\begin{array}{c}
   \delta \rho \\
   \delta  \epsilon \\
   \delta {u}^i \\
   \delta E_{R(\nu)} \\
   \delta {u}_{R(\nu)}^i \end{array}\right)
=\left(\begin{array}{c}
   \rho^{m + 1} - \rho^m \\
   \epsilon^{m + 1} -  \epsilon^m \\
   ({u}^i)^{m+1} - ({u}^i)^{m} \\
   E_{R(\nu)}^{m + 1} - E_{R(\nu)}^m \\
   \left[{u}_{R(\nu)}^i\right]^{m+1} - \left[{u}_{R(\nu)}^i\right]^m \end{array}\right)
~.
\end{eqnarray}
Plugging the expanded form of each variable into equation (\ref{eqn:iterate}), 
we get the following set of equations for the primitive fields $\delta P^a$
\begin{equation}
\sum_a \left(\frac{\partial\mathbf{U}^m}{\partial P^a} - 
             \Delta t~\frac{\partial\mathbf{S_r}^m}{\partial P^a}\right)
       \delta P^a
       = \mathbf{U}^* - \left(\mathbf{U}^m - \Delta t ~\mathbf{S_r}^m\right) ~,
\end{equation}
which is in linear matrix form ${\bf A}{\bf x} = {\bf b}$ with Jacobian
\begin{equation}
A_{ba} = \left(\frac{\partial U^b}{\partial P^a} - \Delta t ~\frac{\partial S_r^b}{\partial P^a}\right) ~,
\label{eqn:jacmatrix}
\end{equation}
or more explicitly
\begin{equation}
{\bf A} = \left( \begin{array}{ccccc} 
  \frac{\partial D}{\partial \rho} 
       & 0 
       & \frac{\partial D}{\partial {u}^i} 
       & 0 
       & 0 \\ 
  \frac{\partial {\cal E}}{\partial \rho}                     + \Delta t \sqrt{-g}\frac{\partial G_0}{\partial \rho} 
       & \frac{\partial {\cal E}}{\partial \epsilon}          + \Delta t \sqrt{-g}\frac{\partial G_0}{\partial \epsilon} 
       & \frac{\partial {\cal E}}{\partial {u}^i}   + \Delta t \sqrt{-g}\frac{\partial G_0}{\partial {u}^i} 
       &                                                        \Delta t \sqrt{-g}\frac{\partial G_0}{\partial E_R} 
       &                                                        \Delta t \sqrt{-g}\frac{\partial G_0}{\partial {u}_R^i} \\
  \frac{\partial {\cal S}_j}{\partial \rho}                   - \Delta t \sqrt{-g}\frac{\partial G_j}{\partial \rho} 
       & \frac{\partial {\cal S}_j}{\partial \epsilon}        - \Delta t \sqrt{-g}\frac{\partial G_j}{\partial \epsilon} 
       & \frac{\partial {\cal S}_j}{\partial {u}^i} - \Delta t \sqrt{-g}\frac{\partial G_j}{\partial {u}^i} 
       &                                                      - \Delta t \sqrt{-g}\frac{\partial G_j}{\partial E_R} 
       &                                                      - \Delta t \sqrt{-g}\frac{\partial G_j}{\partial {u}_R^i} \\
                                                               - \Delta t \sqrt{-g}\frac{\partial G_0}{\partial \rho} 
       &                                                       - \Delta t \sqrt{-g}\frac{\partial G_0}{\partial \epsilon} 
       &                                                       - \Delta t \sqrt{-g}\frac{\partial G_0}{\partial {u}^i} 
       & \frac{\partial {\cal R}}{\partial E_R}               - \Delta t \sqrt{-g}\frac{\partial G_0}{\partial E_R} 
       & \frac{\partial {\cal R}}{\partial {u}_R^i} - \Delta t \sqrt{-g}\frac{\partial G_0}{\partial {u}_R^i} \\
                                                                \Delta t \sqrt{-g}\frac{\partial G_j}{\partial \rho} 
       &                                                        \Delta t \sqrt{-g}\frac{\partial G_j}{\partial \epsilon} 
       &                                                        \Delta t \sqrt{-g}\frac{\partial G_j}{\partial {u}^i} 
       & \frac{\partial {\cal R}_j}{\partial E_R}             + \Delta t \sqrt{-g}\frac{\partial G_j}{\partial E_R} 
       & \frac{\partial {\cal R}_j}{\partial{u}_R^i}+ \Delta t \sqrt{-g}\frac{\partial G_j}{\partial {u}_R^i}
\end{array} \right) ~,
\label{eqn:A}
\end{equation}
and
\begin{equation}
{\bf b} = \mathbf{U}^* - (\mathbf{U}^m - \Delta t ~\mathbf{S_r}^m)
        = \left( \begin{array}{c} 
                 D^* - D^m \\ 
                 {\cal E}^* - {\cal E}^m - \Delta t \sqrt{-g} G_0^m \\ 
                 {\cal S}_j^* - {\cal S}_j^m + \Delta t \sqrt{-g} G_j^m \\ 
                 {\cal R}^*_{(\nu)} - {\cal R}^m_{(\nu)} + \Delta t \sqrt{-g} G_{0(\nu)}^m \\ 
                 {\cal R}_{j(\nu)}^* - {\cal R}_{j(\nu)}^m - \Delta t \sqrt{-g} G_{j(\nu)}^m \end{array} \right) ~.
\end{equation}

Notice that ${\bf A}$ is really a matrix of size $(5+4N_B)\times(5+4N_B)$, and ${\bf x}$ and ${\bf b}$ are  
$(5+4N_B)$-dimensional vectors, where $N_B$ is the number of frequency bins; we have simply condensed 
the notation by representing each 3-vector and all spectral components in ${\bf A}$, ${\bf x}$, and ${\bf b}$ as single entries.  
The linear system only includes terms known at iteration $m$.  
From these we can solve for the vector of unknown primitives at iteration $m+1$, ${\bf P}^{m+1} = {\bf P}^m + {\bf x}$, 
by inverting the matrix ${\bf A}$, solving for ${\bf x}$, then repeating until $\delta{\bf P}/{\bf P}$ converges
to a specified tolerance, which we typically set to $<10^{-5}$ for all primitive fields. 

In addition to this analytic approach we have also developed a numerical method for calculating
the Jacobian matrix that is based on a forward difference approximation to the derivatives,
in which all conserved fields and source terms are evaluated as functions of the primitive iterates.
We have tested both analytic and numerical procedures for calculating the derivatives in (\ref{eqn:A}).  
Both produce consistent results, but we presently use the analytic method as our primary (default) option as it
is faster.
Appendix \ref{sec:derivs} summarizes all of the derivative expressions needed for calculating ${\bf A}$ analytically. 

While this approach is limited to first-order accuracy in time whenever the radiation source terms are stiff (i.e., for large optical depths), it does have the advantages that it: 1) is relatively easy to implement; 2) is stable over a much broader range of parameters than a fully explicit scheme \citep[see][]{fragile14}; and 3) is asymptotic preserving \citep{pareschi01}. In future work, we plan to look into implementing higher-order implicit-explicit (IMEX) schemes. This strategy for solving the radiation source terms clearly also accomplishes the primitive 
inversion step, since we ultimately end up with the set of primitives $\mathbf{P}$ at the new timestep $n+1$.  
The difference with radiation coupling is that it introduces timestep-dependent elements into the Jacobian
matrix so care must be taken in how, when, and how often this operation is performed in the time sequence.

\section{Test Problems}
\label{sec:tests}

For all problems, the Courant factor is set to $k_\mathrm{CFL} = 0.1$, except for the radiation shock tube cases ($k_\mathrm{CFL} = 0.3$) and the free-streaming wave front test ($k_\mathrm{CFL} = 0.2$). The primitive solver tolerance is set to $10^{-5}$, except for the radiation shock tubes, which use $10^{-8}$. All calculations use the \citet{levermore84} closure functions (\ref{eqn:chi}) and (\ref{eqn:xi}). Unless otherwise noted, the radiation energy is plotted in the Eulerian (lab) frame.

\subsection{Free-Streaming Wave Front}
\label{sec:freestream}

\begin{figure}
\begin{center}
\includegraphics[width=0.6\textwidth]{\figpath/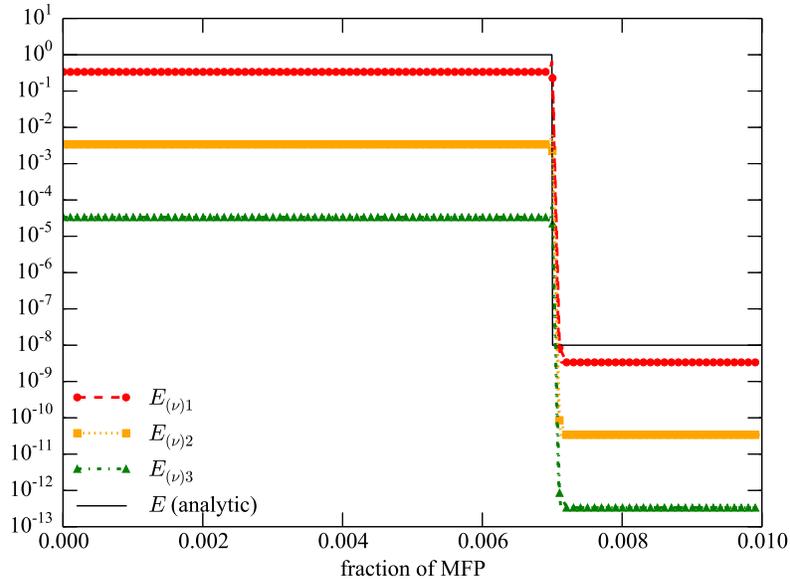}
\caption{
Binned spectral energy densities, $E_{(\nu)n}$, and the analytic frequency-integrated energy density, $E$, for the free-streaming wave problem. Note that the units of $E_{(\nu)n}$ are energy density per energy, while the units of $E$ are just energy density. The horizontal axis in in units of the photon mean free path (MFP) for this problem.
}
\label{fig:stream}
\end{center}
\end{figure}

An important advantage of the $\bf{M}_1$ closure method is its ability to capture the free streaming
limit much more accurately than an energy diffusion scheme, even with the steepest flux limiters.
So this is an appropriate problem with which to begin testing as it is also the
simplest to validate. For this test, we set a background gas with $\rho = 1$ g cm$^{-3}$, $T = 10^4$ K, and $\kappa = 1$ cm$^2$ g$^{-1}$. We then heat the left boundary
to a temperature of $10^6$ K and set the initial radiation energy to $E=a_r T^4$ and
radiation flux to $F=0.999 cE$. 
The entire grid length is fixed at 0.01 of a mean free path so the medium remains optically thin to
photons throughout their propagation history. We assume a constant opacity in temperature,
density, and frequency, but we nevertheless run this test with multiple (3) frequency groups,
assigning spectral energies $E_{(\nu)n} = E/(h\delta\nu_n)/N_b$ 
and spectral
fluxes $F_{(\nu)n} = c E_{(\nu)n}$, where $\delta\nu_n$ is the width of bin $n$ and $N_b=3$ is the number of frequency bins. The bins are spaced logarithmically from $10^{-2}$ to $10^4$ eV. Figure \ref{fig:stream} plots the three spectral energy densities together with
the analytic solution for the total energy density (normalized such that $E = \int E_{(\nu)} d\nu = 1$ behind the wave front) after
the wave front has traveled roughly 70\% of the grid length. 
Note that the spectral energy densities, $E_{(\nu)n}$, are in units of energy density per energy [$E/(h\delta\nu)$], so they are shifted in amplitude from the analytic solution, which is just an energy density, by the different spectral bin widths (and number of bins). We intentionally plotted it this way to separate out the energy density profiles for clarity. It is easy to confirm, however, that $E_{(\nu)1} \delta\nu_1 + E_{(\nu)2} \delta\nu_2 + E_{(\nu)3} \delta\nu_3 \approx E$.
We observe a slight overshoot in $E_{(\nu)n}$ of roughly 10-15\% at the front edge of the wave, but 
otherwise the numerical and analytic solutions agree quite nicely.
The average relative error, $\sum_i \vert a_i - A_i \vert/(N A_i)$, where $a_i$ and $A_i$ are the numerical and analytic solutions, respectively, of $E$ is $<8\times10^{-3}$ for $N=1000$ zones, and converges at
a rate slightly faster than first order, as expected for problems with sharp discontinuities.

\subsection{Diffusive Point Source}
\label{sec:diffusion}

In the opposite optically thick limit,
\citet{pons00} proposed an analytically tractable problem describing the propagation of
a single point source in a strongly diffusive medium.
The medium is endowed with zero absorptivity but very high scattering opacity in each frequency bin $\kappa^s_{(\nu)}$.
For a sufficiently opaque, spherically symmetric medium, the lab-frame energy and flux evolve as a function
of radius ($r$) and time ($t$) according to
\begin{equation}
E(r,t) = \left(\frac{k_s}{t}\right)^{3/2} \exp\left(\frac{-3 k_s r^2}{4 c t}\right)  ~,
\label{eqn:diff_ene}
\end{equation}
\begin{equation}
F^r(r,t) = \frac{r}{2 t} E ~.
\label{eqn:diff_flux}
\end{equation}

\begin{figure}
\begin{center}
\includegraphics[width=0.6\textwidth]{\figpath/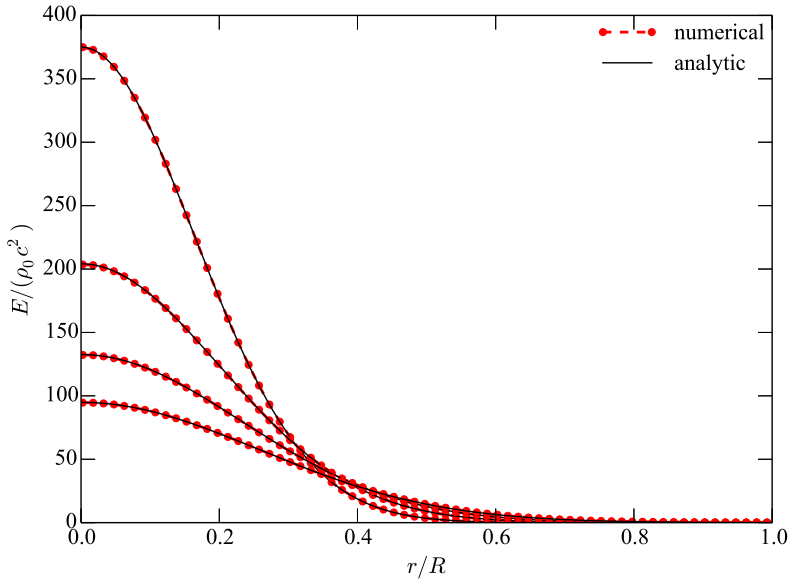}
\caption{
Frequency-integrated radiation energy density for the diffusive point source test,
calculated on a grid of 200 zones at times $t=200$, 300, 400, and 500.
The numerical solution converges at second order toward the analytic solution.
}
\label{fig:diffusion}
\end{center}
\end{figure}

The tests presented here fix the grid length to $R=2 \times 10^9$ cm, the gas density to $\rho_0 = 9\times10^{14}$ g/cm$^3$,
and the gas temperature to $5\times10^6$ eV. Specification of the Peclet number $P_e = \kappa_s\Delta \ell$,
where $\Delta \ell$ is a characteristic scale, defines the scattering opacity.
We tie the length scale to a small fraction of the grid length $\Delta \ell=R/50$ and set $Pe=100$, safely
within the strong scattering limit.
All calculations are initialized at $t_0=200R/c$ and, for the convergence studies, run out 
to $t=1.5 t_0$, enough time for the solutions to decay to roughly half of their initial peak energies.
Similar to the streaming test, these problems are run with three logarithmically spaced
spectral bins ranging from $10^4$ to $10^8$ eV and initialized with spectral energy densities $E_{(\nu)}$
and fluxes $F^r_{(\nu)}$ such that the frequency-integrated lab frame energy density and flux 
equate to equations ($\ref{eqn:diff_ene}$) and ($\ref{eqn:diff_flux}$).
Results for $E(r)$ on a grid with $N=200$ zones are plotted in Figure \ref{fig:diffusion}. We find average (maximum) errors of $E$ of $2.3\times10^{-3}$ ($8.7\times10^{-3}$) and
$5.9\times10^{-4}$ ($2.6\times10^{-3}$) on grids of $N=100$ and 200 zones, respectively, consistent
with a second order convergence rate.

\subsection{Picket Fence}
\label{sec:picket}

Analytic benchmark solutions for non-equilibrium radiative transfer are scarce, and this is
especially true for multi-frequency general relativistic transfer. 
Hence we occasionally resort, as we do in this section,
to the Newtonian literature and limit. As we have emphasized, the radiative transfer algorithms
in {\em Cosmos++} are adapted to work in both Newtonian and general relativistic regimes,
and because of the covariant nature of the formalism, much of the coding is shared.  
In fact the only major difference between the two is the primitive inversion scheme, 
which is not needed for the Newtonian limit. As a result Newtonian problems will exercise
much of the general relativistic coding. 

One particularly interesting Newtonian problem is the picket-fence proposed by \cite{su99}. This
test provides a semi-analytic solution for non-grey, two-temperature, non-equilibrium
radiative transfer and diffusion, with the caveats that the opacity must be independent of temperature and
the material specific heat must be proportional to the cube of the temperature, i.e., $C_v = \alpha T^3$.
This problem is initialized with a cold, purely 
absorbing medium, then heated by an extended isotropic radiation source that is
a function of space, time and frequency $S(x,t,\nu)$. The source is actually
constant over space and time, but active only for a finite duration and over a finite region of space.
Hydrodynamic motion (other than thermal coupling) is ignored.

\begin{figure}
\begin{center}
\includegraphics[width=0.6\textwidth]{\figpath/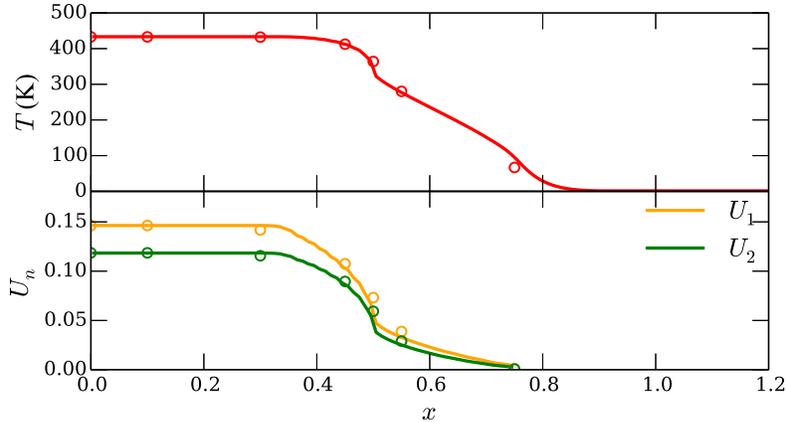}
\caption{
Temperature (top panel) and frequency-integrated radiation energy densities (bottom panel) for the picket-fence test Case $\mathscr{B}$ from \citet{su99} at $\tau = 0.3$ run
on a grid of 300 zones.
The analytic, transport solutions are included as symbols.
}
\label{fig:picket}
\end{center}
\end{figure}

The cold medium is initialized with unit density,
unit temperature, and specific heat constant $\alpha=4 a_r/\epsilon$, where $a_r$ is the
radiation constant and $\epsilon=1$.
The multi-frequency aspect of this test is scripted in the opacity, which
is assumed to take one of two values, $\kappa(\nu) = \kappa_n$ where $n=(1,2)$,
across alternating frequency bins. We use 20 bins to cover (logarithmically)
the frequency range $10^{-8}$ to 10 eV$/h$, and set the alternating opacities to 2 and 20.
The radiating source emits at the rate
$S=c a_r \rho \kappa_0 T_0^4$ erg/s/cm$^3$, where $T_0=10^3$ K and $\kappa_0=(\kappa_1 + \kappa_2)/2$ 
is the mean opacity, corresponding to Case $\mathscr{B}$ from \citet{su99}.
The numerical box size is set to three mean free paths and the radiating source is contained
within half a mean free path of the left-most edge of the grid.
Figure \ref{fig:picket} plots the results at $\tau = 0.3$, where $\tau$ is the time measured in units of $\alpha/(4 c a_r \rho \kappa_0)$. The upper panel presents the
gas temperature, while the lower panel shows the radiation energies corresponding to the quantities $U_1$ and $U_2$ 
in the notation of \citet{su99}, which represent
the total integrated energies across each of the two opacity intervals, i.e., $U_n =\int E_{(\nu)n} d\nu_n$. We also include the benchmark transport solutions as tabulated in \citet{su99}. The agreement with the numerical solutions
is better than 10\% throughout the temperature and energy profiles, a fairly good
agreement considering that the $\bf{M}_1$ closure makes different assumptions than the transport model.

\subsection{Homogeneous Radiating Sphere}
\label{subsec:radsphere}

We next consider two variants of the homogeneous radiating sphere test \citep{muller10}.
The basic configuration consists of a static, spherically symmetric, homogeneous, 
and isothermal stellar sphere of radius $R_\star$, which radiates into a surrounding vacuum region.
We assume the dominant interaction process inside the sphere is isotropic
absorption and thermal emission with constant absorption opacity $\kappa^a_{(\nu)}$ and
emissivity $B$. Under such conditions, this problem has the following analytic
solution \citep{smit97}
\begin{equation}
I(r,\mu) = B \left( 1 - e^{-\kappa s(r,\mu)}\right) ~,
\label{eq:radsphere_soln}
\end{equation}
where
\begin{equation}
s(r,\mu) =
    \begin{cases}
       r\mu + R_\star g(r,\mu) & \text{if}\ r < R_\star,    \quad -1 \le \mu \le 1, \\
       2 R_\star g(r,\mu)      & \text{if}\ r \ge R_\star,  \quad \sqrt{1-(R_\star/r)^2} \le \mu \le 1, \\
       0,                  & \text{otherwise,}
    \end{cases}
\end{equation}
\begin{equation}
g(r,\mu) = \sqrt{1 - \left(\frac{r}{R_\star}\right)^2 \left(1-\mu^2\right) } ~,
\end{equation}
and $\mu = \cos \theta$ is the directional cosine, such that this solution is an integral over all directions. The radiation energy and flux are derived via a numerical integration of the first two moments (angular integrals) of (\ref{eq:radsphere_soln}).

\begin{figure}
\includegraphics[width=0.5\textwidth]{\figpath/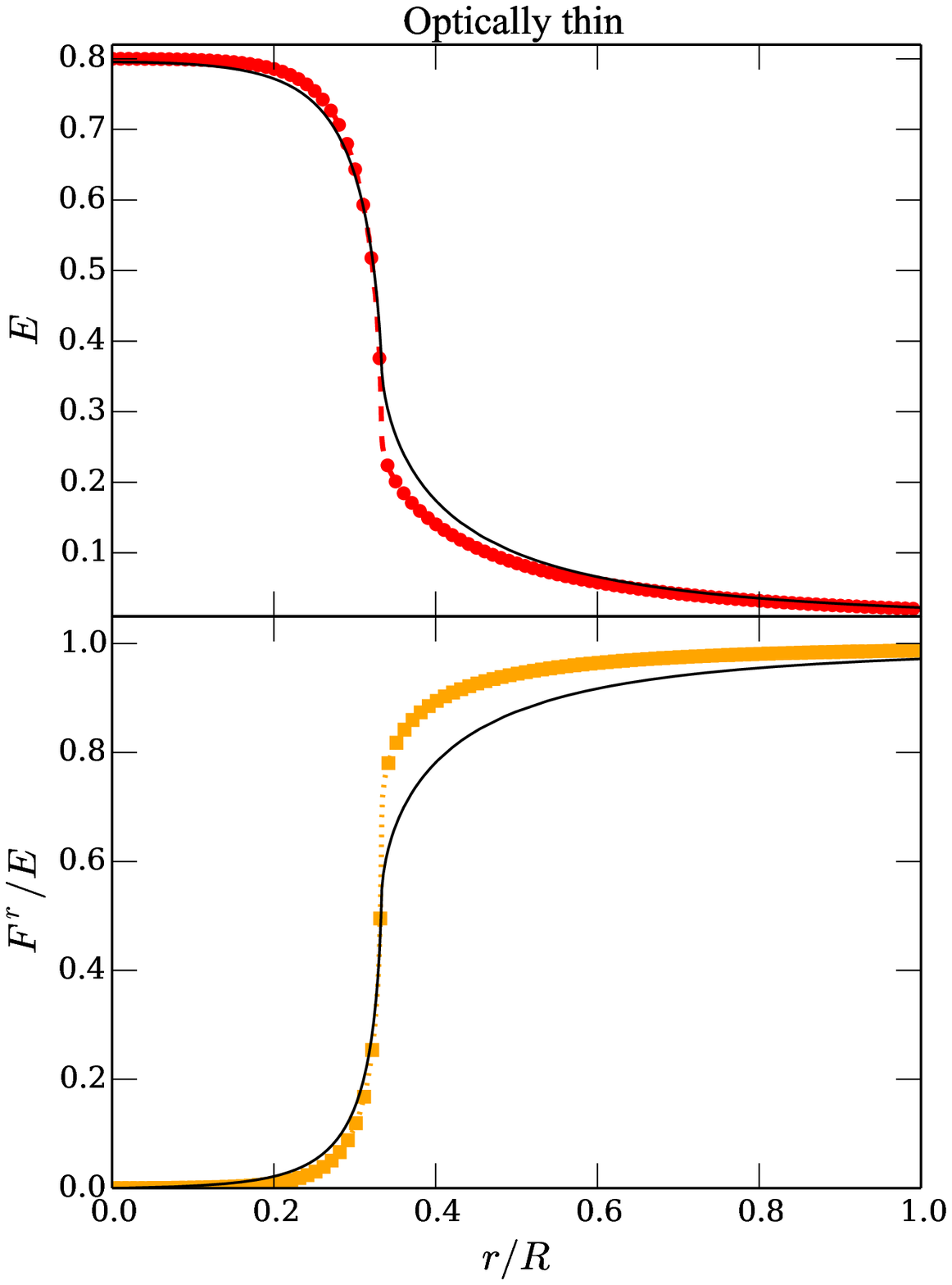}
\includegraphics[width=0.5\textwidth]{\figpath/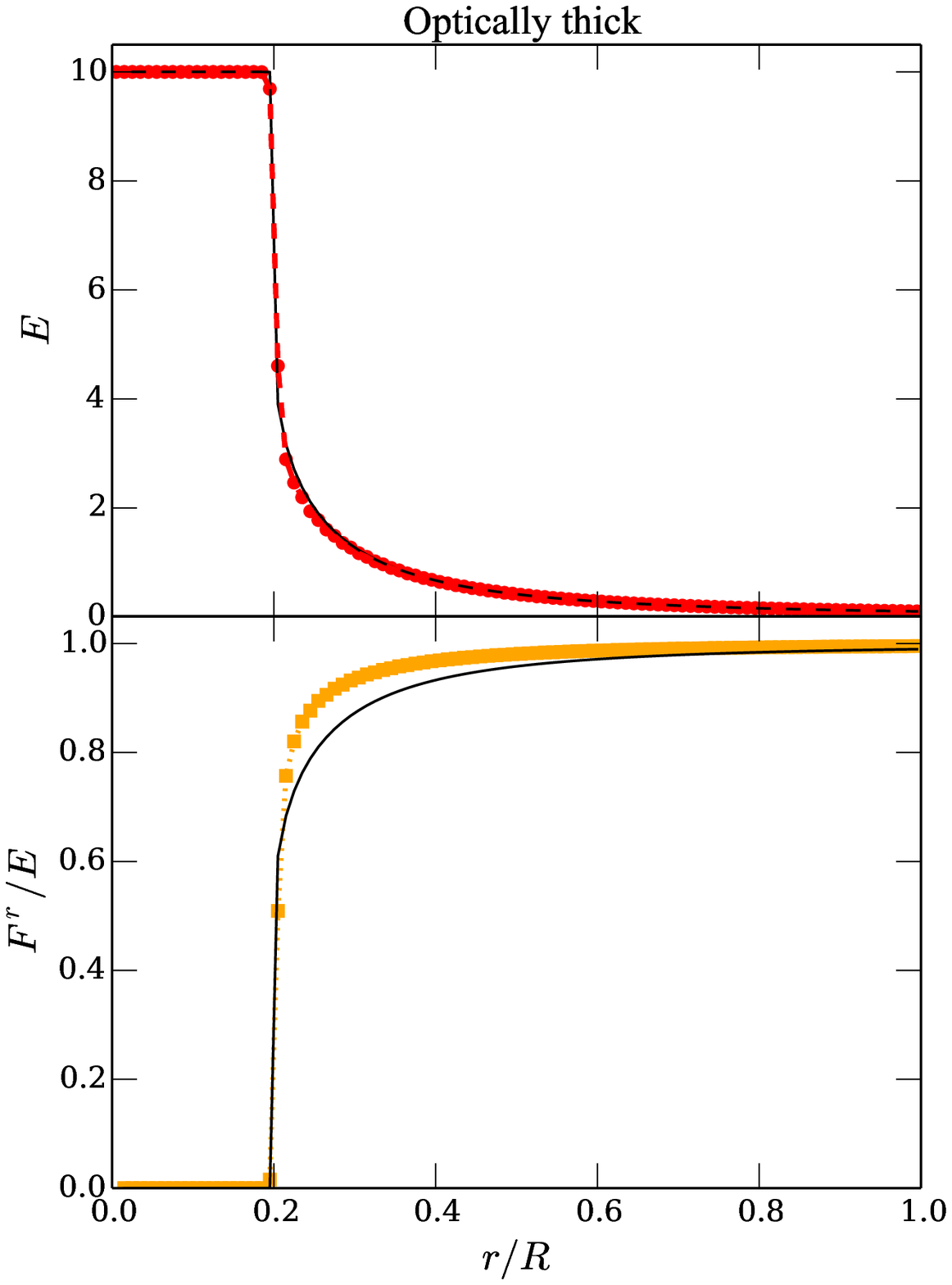} 
\caption{
Steady-state radiation energy densities (top panels) and radial flux ratios (bottom panels) 
for the homogeneous radiating sphere tests.
The left (right) panels are the optically thinner (thicker) solutions. Symbols show the numerical solutions, while the thin black lines give the analytic ones.
}
\label{fig:radsphere}
\end{figure}

We set the sphere radius to $R_\star=10$ km
and initialize the interior with a constant density $\rho_0=9\times10^{14}$ g cm$^{-3}$,
a constant temperature with the parametrization $T_0 = (B/a_r)^{1/4}$, and
opacity $\kappa = P_e/\Delta r$ where $\Delta r = R/N_r$ is the cell size, $R$ is
the grid domain length, $N_r$ is the number of grid cells, and $P_e$ is the Peclet number. 
The exterior background density is fixed at $10^{-10} \rho_0$.
We consider two parameter sets representing different optical regimes:
an optically thinner case \citep{smit97} with $N_r=1000$, $R=3 R_\star$, $B=0.8$, and $P_e=0.015$; and a highly thick
case \citep{abdikamalov12} with $N_r=100$, $R=5 R_\star$, $B=10$, and $P_e=12.5$.
The steady-state solutions for the radiation energies and radial flux-to-energy ratios ($F^r/E$)
are shown in Figure \ref{fig:radsphere} together with the 
corresponding analytic solutions. 
The top panels plot the radiation energies, while the bottom panels are the radial flux ratios.
The left (right) panels are the optically thinner (thicker) solutions.
The qualitative behavior and results compare well to the solutions. We point out, as have previous authors \citep{smit97, oconnor15}, that these
tests are particularly sensitive to the closure relation which helps explain
the deviations observed near the stellar surface. That our numerical methods can
handle the discontinuities near the surface and match the asymptotic behavior extremely
well is encouraging.

\subsection{Doppler Frequency Shift}
\label{subsec:test_redshift_dop}

The frequency coupling terms from Section \ref{subsec:doppler} are tested using
the same series of calculations as performed by \citet{muller10,oconnor15,kuroda16}. These tests
involve the propagation of radiation from a homogeneous radiating sphere, similar to
Section \ref{subsec:radsphere} except that in addition to isotropic 
absorption and emission, a sharp velocity profile is added outside the sphere to mimic an
accretion flow capable of Doppler redshifting the outgoing radiation. 

For all of these tests we set the stellar radius to $R_\star=10$ km, 
and impose a uniform interior density of $\rho_0=9\times10^{14}$ g cm$^{-3}$ and temperature of 5 MeV$/k$.
The outer radius of the computational box is fixed at $R=800$ km, and in order to resolve
both the radiating sphere and Doppler velocity features we use 1080 cells along the
radial direction resulting in a grid resolution of $\Delta r = 0.74$ km.
We choose to run these tests with neutrinos, rather than photons, and 
with both 15 and 25 frequency bins spanning the range 1-50 MeV$/h$.
The stellar opacity is made sufficiently thick by setting the Pecklet number to unity
over the scale of a single zone.
For the velocity profile we use
\begin{equation}
v(r) =
    \begin{cases}
       0,                                                         & r  \le 70~\text{km}, \\
       -0.2 c\left(\frac{r - 70~\text{km}}{10~\text{km}}\right),  & 70~\text{km} \le r \le 80~\text{km},  \\
       -0.2 c\left(\frac{80~\text{km}}{r}\right)^2,               & r  \ge 80~\text{km} ~.
    \end{cases}
\label{eqn:dop_vel}
\end{equation}

In this section we consider tests of just the Doppler redshift arising from radiation
streaming through the infalling velocity profile of
equation (\ref{eqn:dop_vel}) without a gravitational potential.
However, the same set of analytic solutions are equally applicable to
cases with a redshifting potential as we will see in the next section.
Variations in the luminosity arise in the free-streaming limit when there
are nonzero velocities or potentials, and satisfy the following  analytic solution
\begin{equation}
{\cal L}(r) \propto \frac{w}{\alpha} \frac{1-v}{1+v} \alpha r^2 g^{rr} \int F_r d\nu \propto \frac{w}{\alpha} \frac{1-v}{1+v} ~,
\label{eqn:lumin}
\end{equation}
where $w$ is the Lorentz factor, $\alpha(r)$ is the lapse function, and the quantity $\alpha r^2 g^{rr} \int F_r d\nu$ is the luminosity measured by an
Eulerian observer and should be constant far from the star. 
This implies that the mean neutrino energy as measured in the co-moving frame can be
calculated as a function of radius
\begin{equation}
h\langle\nu\rangle(r) = \frac{\alpha(R_\star)}{\alpha(r)} \frac{h\langle\nu_\star\rangle}{w (1+ v(r))}
\end{equation}
for both Doppler and gravitational redshifts provided the
fluid velocity is zero at the star surface
and $h\langle\nu_\star\rangle = 15.76$ MeV is the mean neutrino energy at the stellar surface.
Because of the high stellar opacity we can assume any
escaping radiation originates from the stellar surface $r=R_\star$.

\begin{figure}
\includegraphics[width=0.5\textwidth]{\figpath/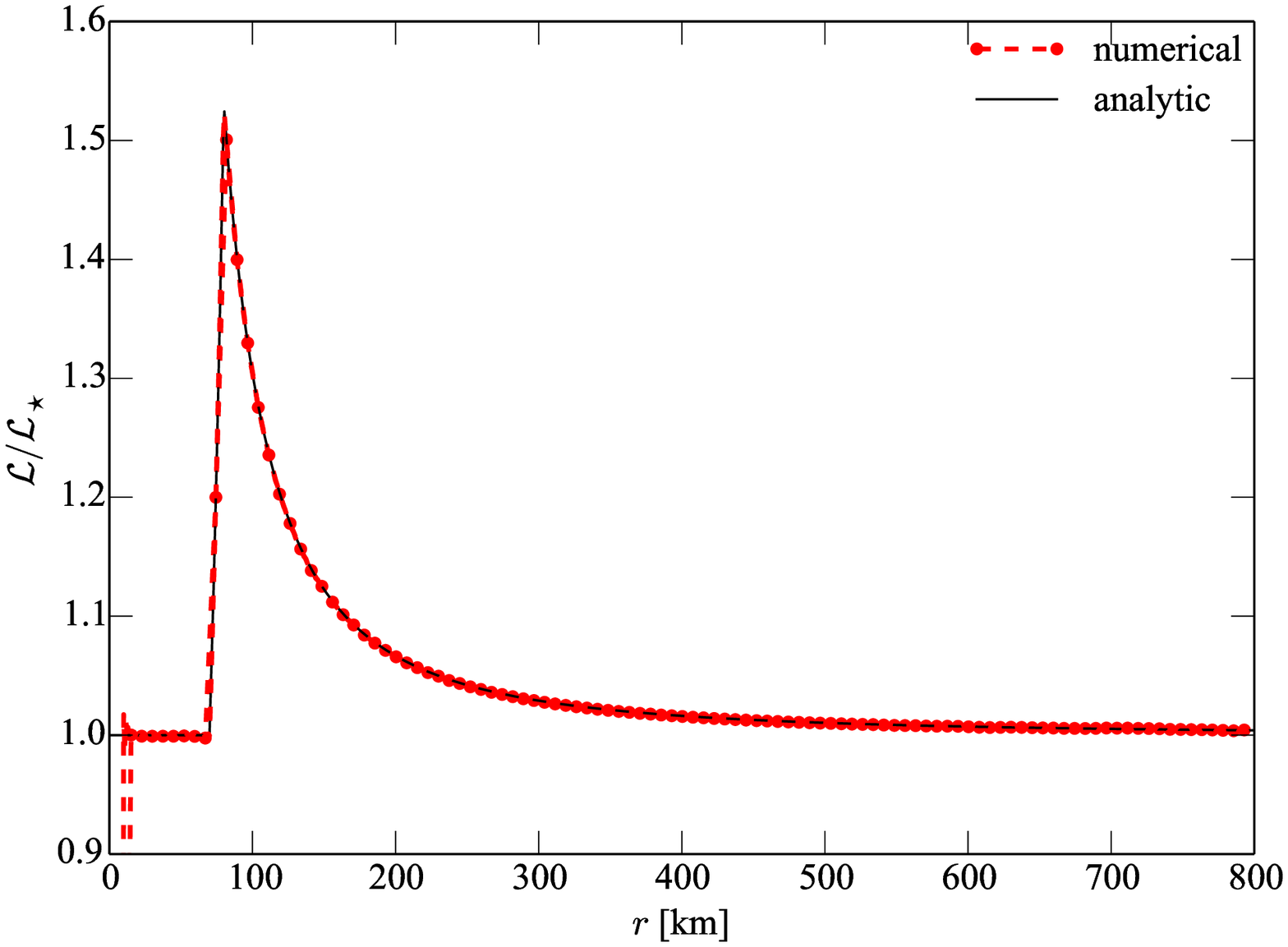}
\includegraphics[width=0.5\textwidth]{\figpath/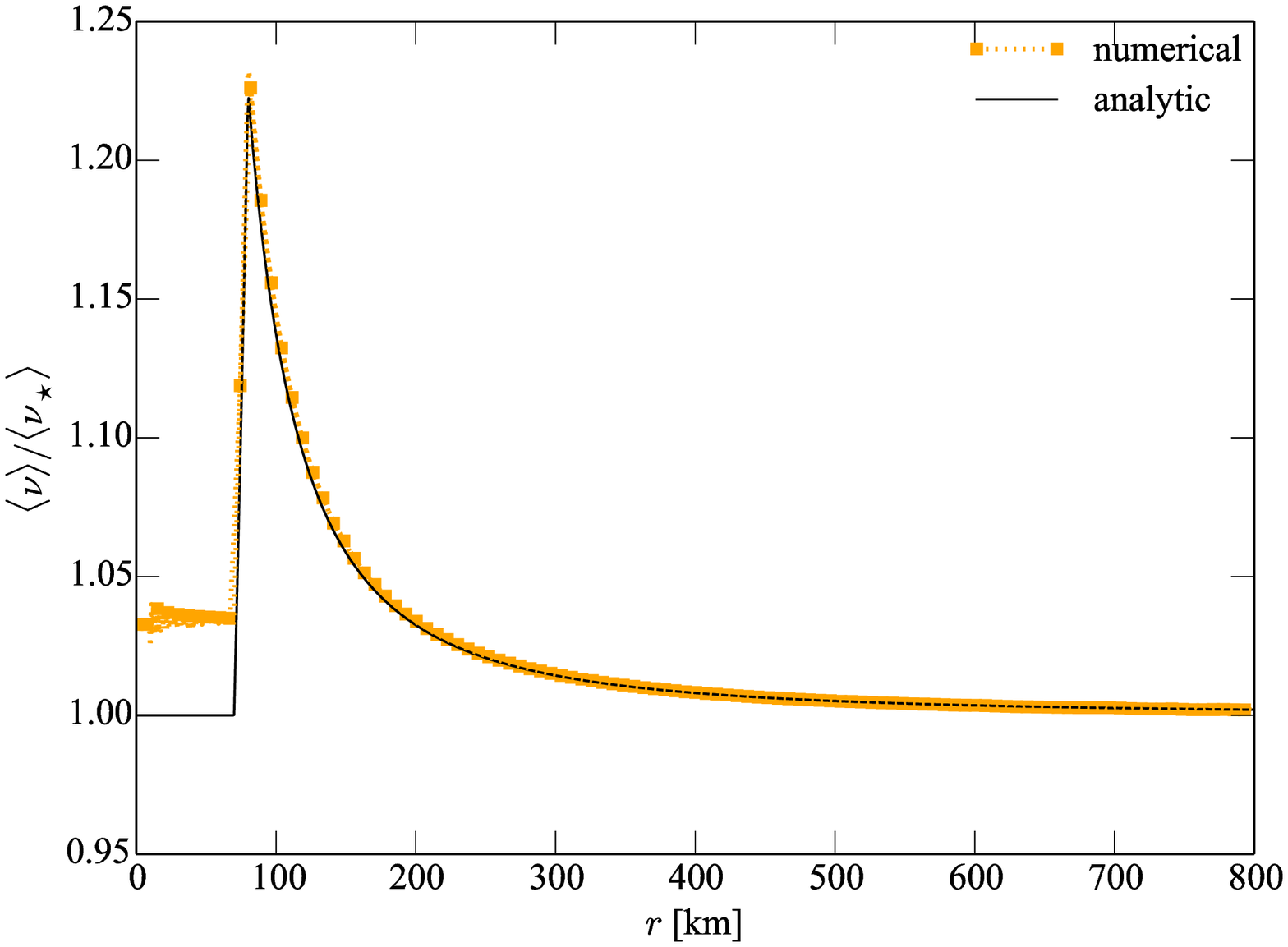} 
\caption{
Steady-state luminosity (left) and mean co-moving neutrino
energy (right) as functions of radius for neutrino
streams undergoing Doppler redshifting from a model accreting velocity source.
}
\label{fig:red_dop}
\end{figure}

Figure \ref{fig:red_dop} shows the luminosity profile 
and the mean co-moving frame energy 
as functions of radius calculated with 25 frequency bins
after the radiation achieves steady-state.
The corresponding analytic solutions, also plotted in Figure \ref{fig:red_dop}, match the numerical
results almost exactly everywhere except near the stellar surface where the maximum error in the average photon
energy (in the right plot) plateaus at about 3\%. However, we point out that the width of the frequency bins 
near the emission frequency ($\nu_\star$) is $\delta\nu_n \approx 2$ MeV$/h$, or about $0.13 \nu_\star$.
Hence the observed error is within the uncertainty of interpolation between frequency bins,
and convergences to zero with increasing frequency resolution, a fact that we have confirmed by
running this identical test with a smaller number of frequency bins (15).
We note that
we importantly observe between first and second order convergence
in matching the mean energies at the peak and in the
region between the stellar surface and the velocity discontinuity.
In particular the maximum errors with 15 (25) frequency bins
in the near-surface plateau and Doppler peak regions are 
5.3\% (3.0\%) and 1.6\% (0.73\%), respectively,
corresponding to a convergence rate in frequency of about 1.5.

\subsection{Gravitational Redshift}
\label{subsec:test_redshift_grav}

The problem from the previous section can readily be amended to test gravitational redshifting, too, which we do now.
Aside from introducing
a self-gravitating potential, all of the problem parameters and the configuration are identical to
those specified previously.
The potential ($\phi < 0$) is calculated by solving the Newtonian
Poisson equation for the uniform density sphere, then assigning 
the spacetime metric in the spherical, Kerr-Schild gauge as
\begin{equation}
ds^2 = -(1 + 2\phi/c^2) dt^2 + (1-2\phi/c^2) dr^2 + r^2 d\Omega^2 ~.
\end{equation}

\begin{figure}
\includegraphics[width=0.5\textwidth]{\figpath/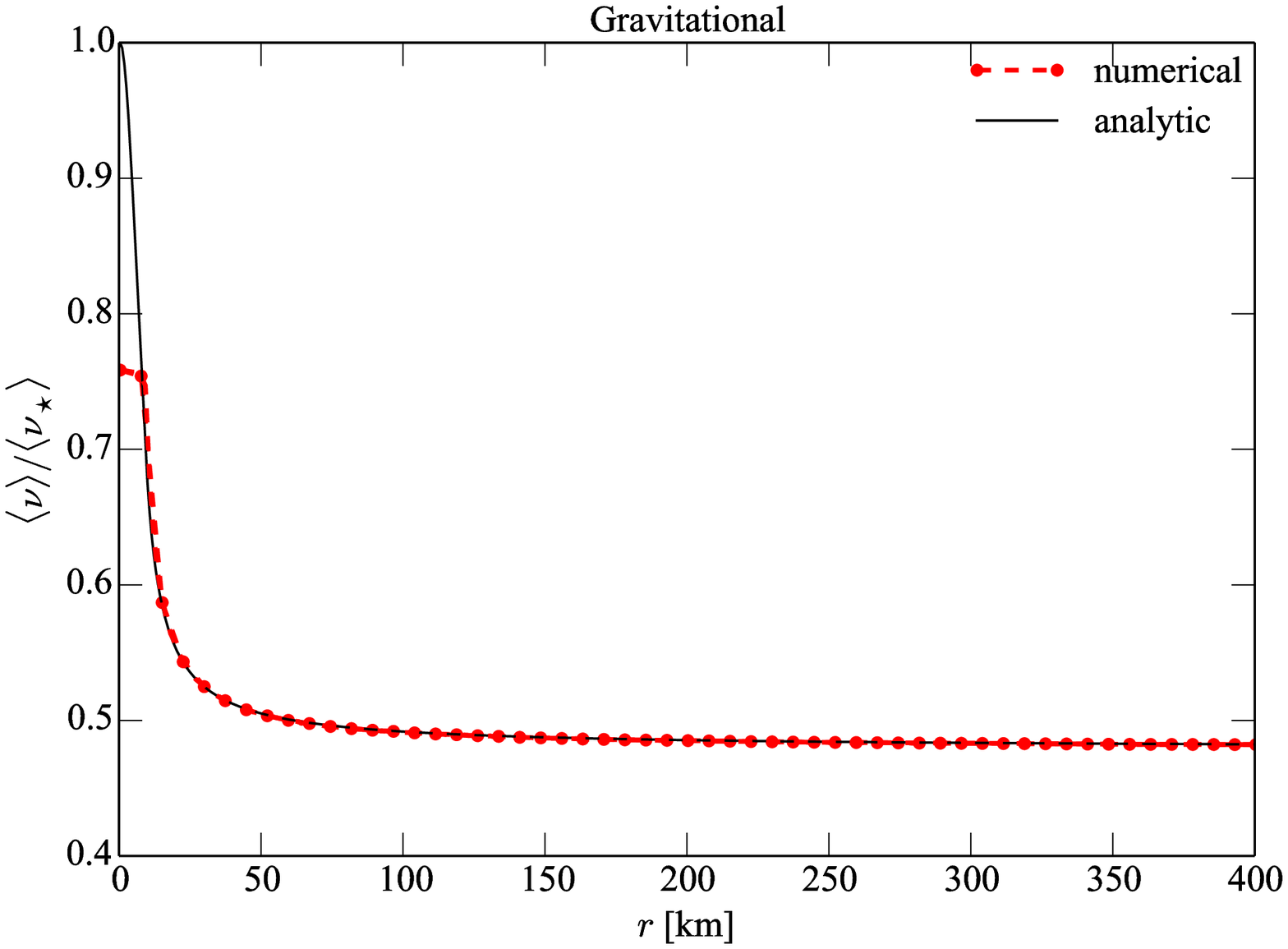}
\includegraphics[width=0.5\textwidth]{\figpath/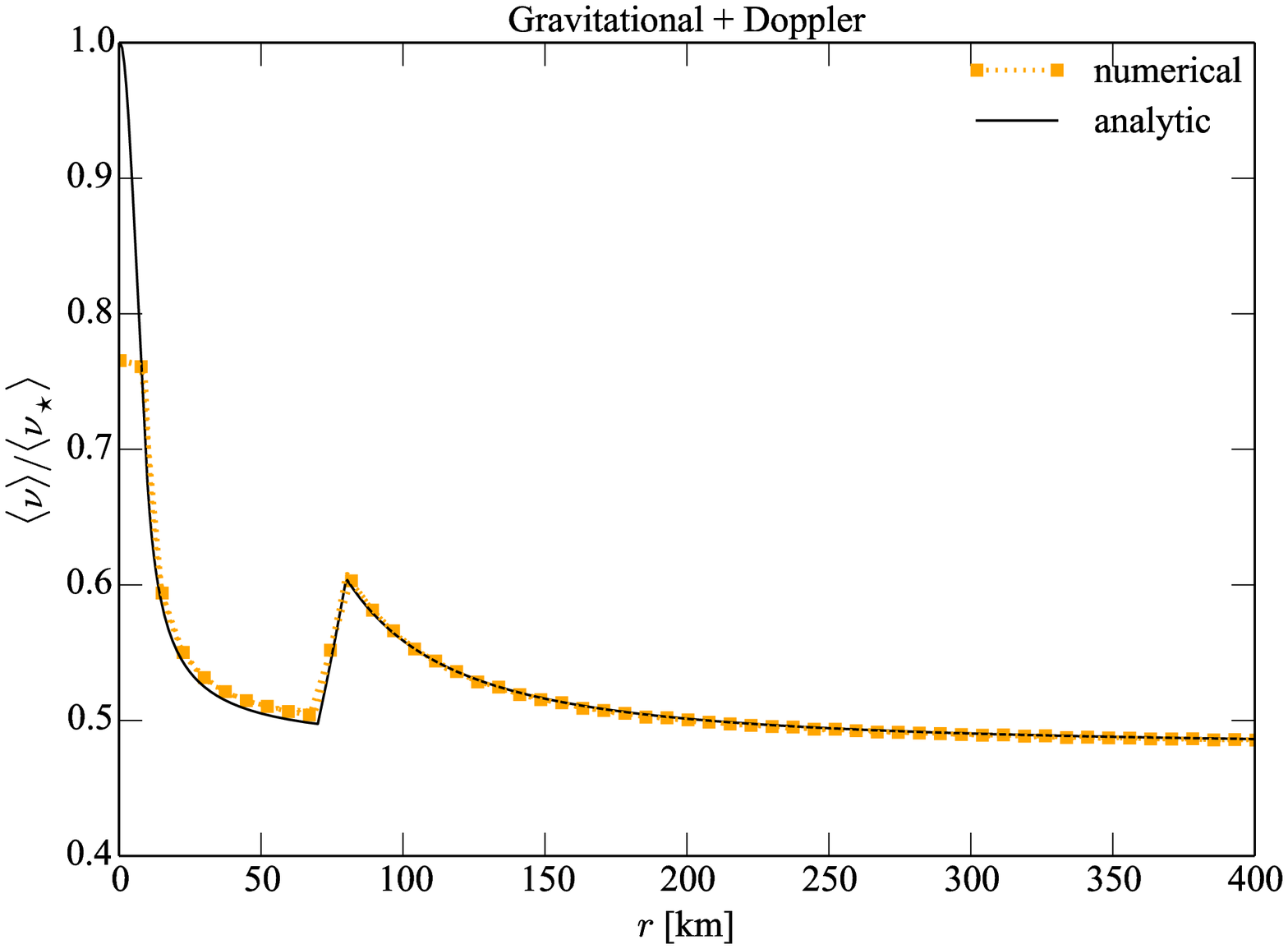} 
\caption{
Mean co-moving neutrino energy as a function of radius for tests of pure gravitational (left) and gravitational plus Doppler (right) redshift.
}
\label{fig:red_grav}
\end{figure}

Figure \ref{fig:red_grav} displays the mean neutrino energy as a function
of radius for two cases: pure gravitational redshift (left) and
a combination of gravitational plus Doppler redshift with the same velocity
profile as specified in Section \ref{subsec:test_redshift_dop} (right).
Following the example in that section, we have run each
scenario with 15 and 25 frequency bins to verify convergence. The solutions
presented in Figure \ref{fig:red_grav} are from the 25-bin tests.
The Doppler peaks agree with the analytic solution
to better than 1\% for both resolutions. As found for the pure Doppler test, agreement is worst
right in front of the velocity discontinuity. This region is
sensitive to both the spatial and frequency resolutions. Our results with 15 (25) bins
nevertheless agree with the analytic solution with maximum errors of about 3.1\% (1.5\%), converging between
first and second order in frequency.

\subsection{Radiation Shock Tube}
\label{subsec:shock}

\begin{figure}
\includegraphics[width=0.5\textwidth]{\figpath/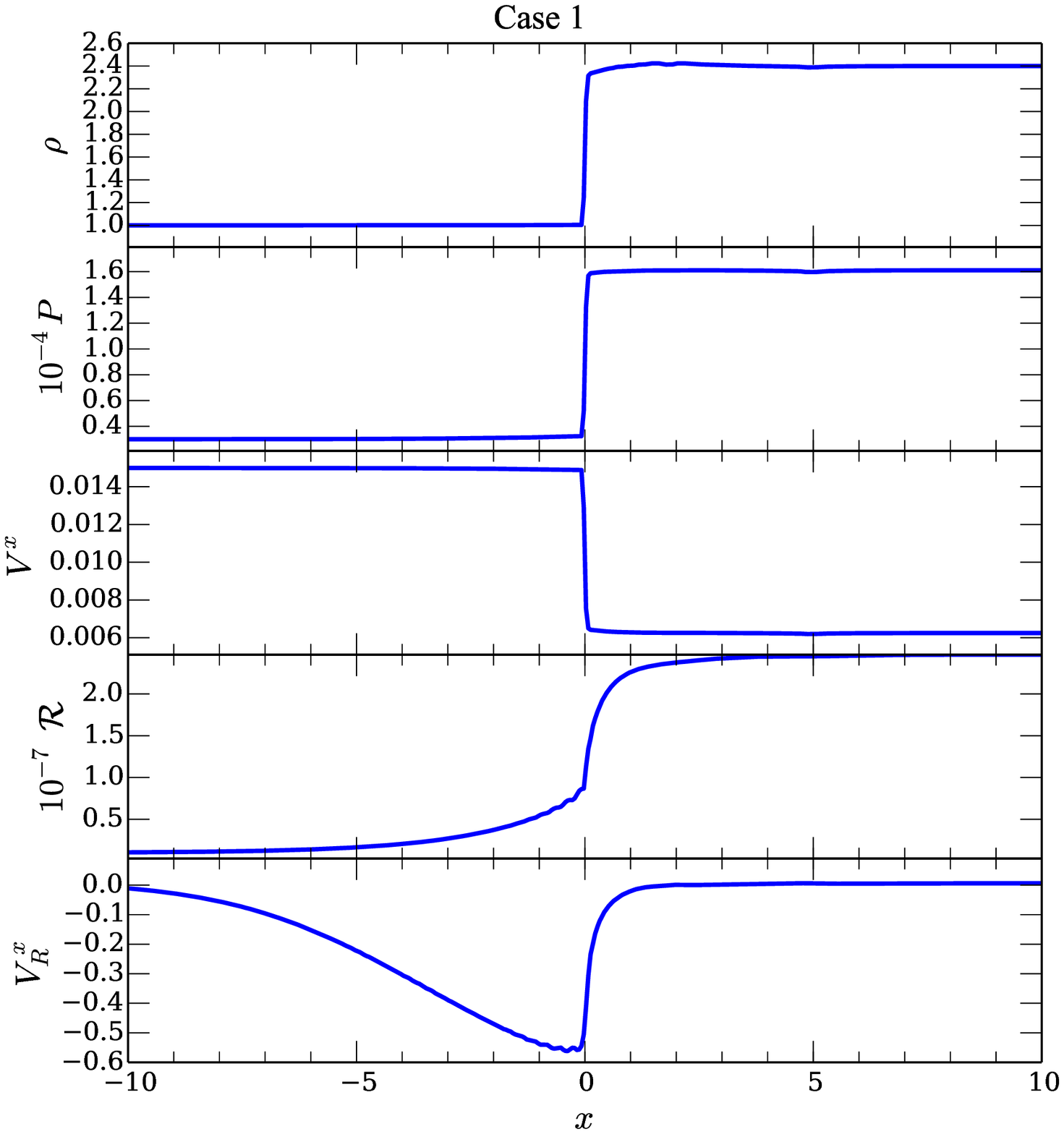}
\includegraphics[width=0.5\textwidth]{\figpath/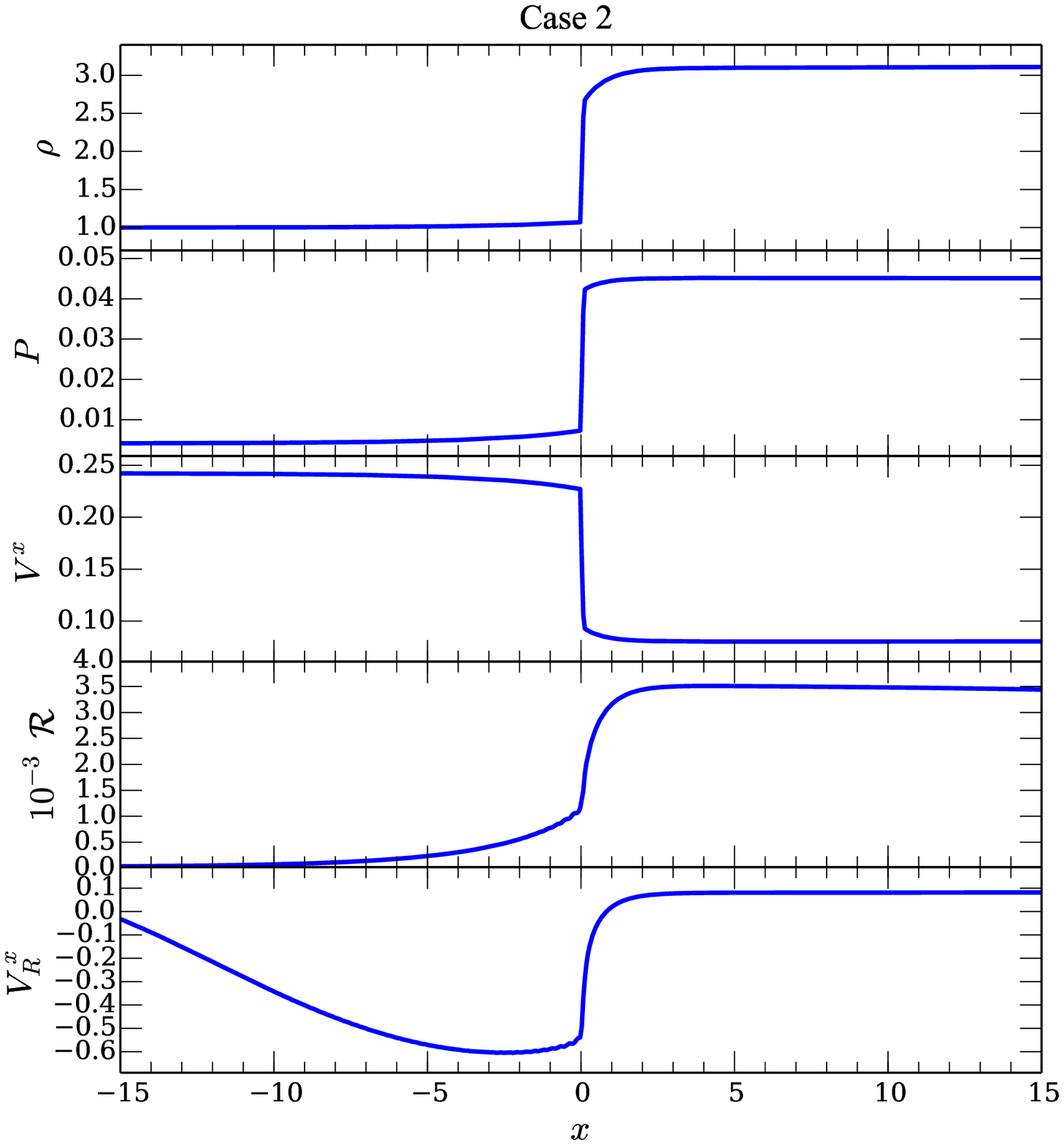} \\
\includegraphics[width=0.5\textwidth]{\figpath/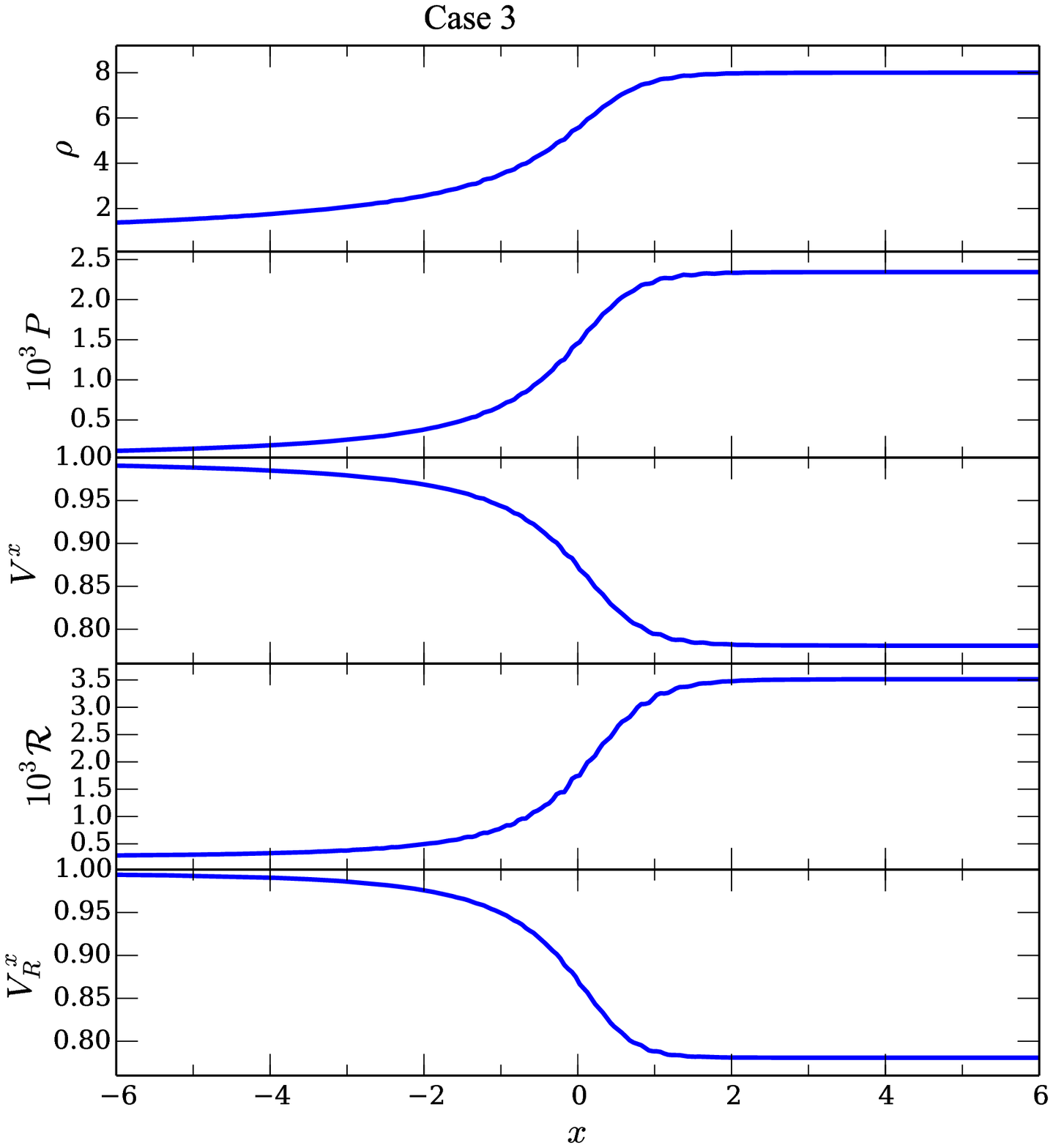}
\includegraphics[width=0.5\textwidth]{\figpath/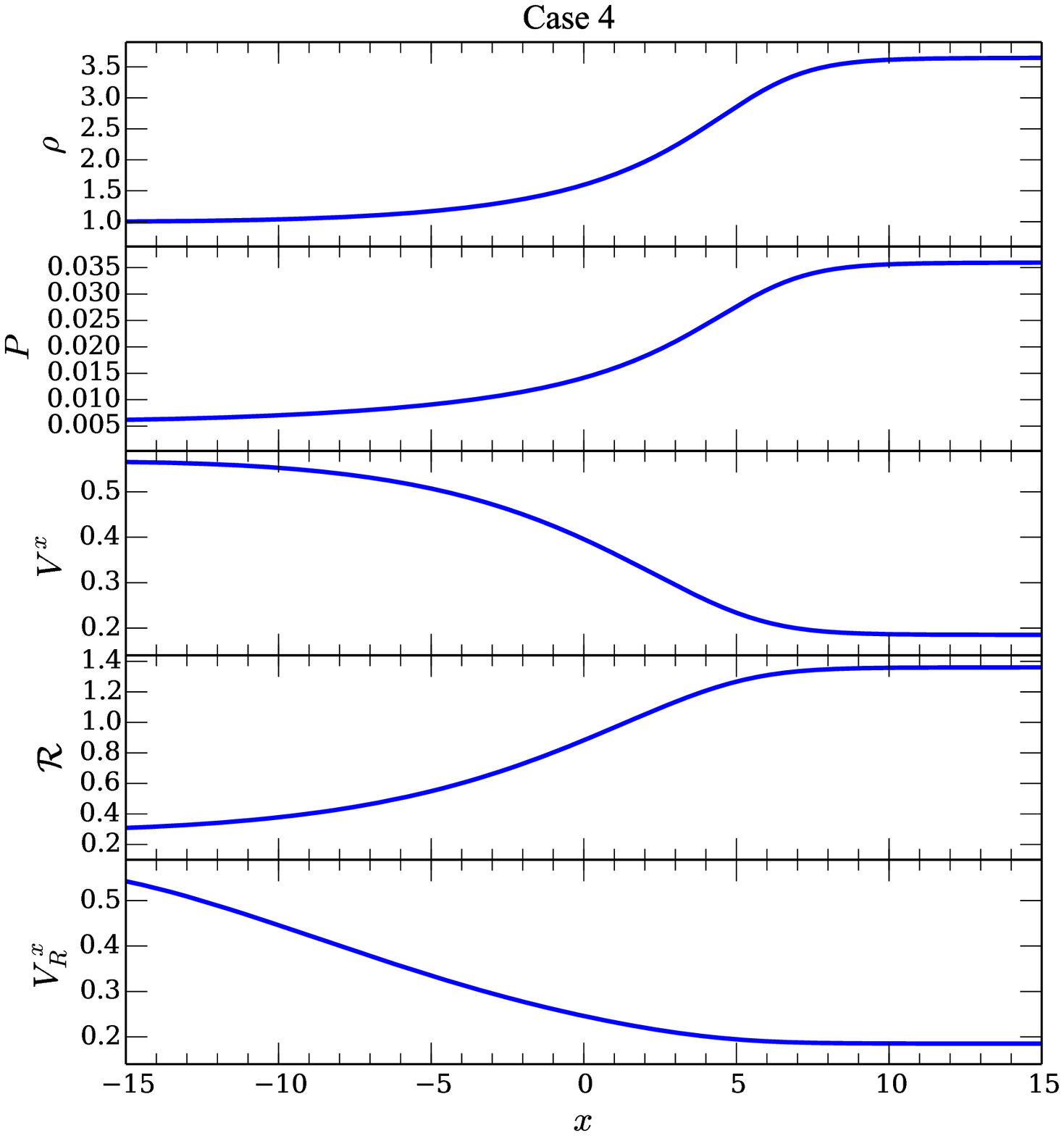}
\caption{
Profiles of gas density, gas pressure, gas velocity, conserved radiation energy, and radiation rest-frame velocity
after $t=300$ for the four radiation shock tube cases.
}
\label{fig:shock}
\end{figure}

Most of the tests we consider in this work do not involve both strong kinetic and thermal coupling between the radiation and matter. To address this, we include the following four radiation shock tube tests first introduced by \citet{farris08}
\citep[see also][]{zanotti11,fragile12,mckinney14}:
a nonrelativistic strong shock (case 1), a relativistic strong shock (case 2),
a relativistic wave (case 3), and a radiation pressure dominated relativistic wave (case 4).
The initial parameters are the same as those in \citet{farris08} and are reproduced in Table \ref{tab:shocktube}. 
These tests are run until $t=300$ on a grid with 800 zones over $x \in [-20,20]$. 
All calculations are run with spectrally uniform energy distributions, $E_{(\nu)n} = E/(h\delta\nu_n)/N_b$, and
eighteen frequency groups with bin edges at $0.05 kT_\mathrm{min}/h$ and $20 kT_\mathrm{max}/h$, where $T_\mathrm{min}$ and $T_\mathrm{max}$ come from the initial states of each case. 
The results are plotted in Figure \ref{fig:shock}, where
we show (in top to bottom order) the gas density, gas pressure, gas velocity, conserved radiation energy,
and radiation rest-frame velocity for each case. We point out that we solve these problems using the $\bf{M}_1$ closure, whereas in most earlier work presenting these tests \citep[e.g.][]{farris08,fragile12,sadowski13,fragile14}, they were solved using the Eddington closure. Not surprisingly, the two closures yield slightly different results, which explains why our new results look different from our previously published ones (we also ran most of the cases to different stop times than before). Only one work that we are aware of, \citet{mckinney14}, has previously published these tests using the $\bf{M}_1$ closure, so our results should be compared with those, and, in fact, we find the agreement to be excellent. 

\begin{deluxetable}{ccccccccccc}
\tablecaption{Radiation Shock Tube Parameters \label{tab:shocktube}}
\tablewidth{0pt}
\tablehead{
\colhead{Case} & $\Gamma$ & $\kappa^a$ &
$\rho_L$       & $P_L$    & $u_L^x$    & $E_L$      &
$\rho_R$       & $P_R$    & $u_R^x$    & $E_R$ 
}
\startdata
1 & 5/3  & 0.4 
  & 1    & $3   \times 10^{-5}$  & $0.0015$              & $1    \times 10^{-8}$   
  & 2.4  & $1.61\times 10^{-4}$  & $6.25\times10^{-3}$   & $2.51 \times 10^{-7}$ \\
2 & 5/3  & 0.2  
  & 1    & $4 \times 10^{-3}$    & 0.25                  & $2    \times 10^{-5}$  
  & 3.11 & $0.04512$             & $0.0804$              & $3.46 \times 10^{-3}$ \\
3 & 2    & 0.3  
  & 1    & $60$                  & 10                    & 2                      
  & 8    & $2.34 \times 10^{3}$  & $1.25$                & $1.14 \times 10^{3}$ \\
4 & 5/3  & 0.08 
  & 1    & $6 \times 10^{-3}$    & 0.69                  & 0.18                   
  & 3.65 & $3.59 \times 10^{-2}$ & $0.189$               & $1.30$ \\
\enddata
\end{deluxetable}

\subsection{Shadow Casting}
\label{sec:cloud}

An important advantage of $\bf{M}_1$ closure over the diffusion (or
isotropic Eddington) approximation is its ability to preserve shadows in the wake of opaque objects.
In a previous paper \citep{fragile14} we demonstrated this ability with our grey (single group)
version of $\bf{M}_1$. In that paper
we considered radiation flow across an opaque spheroidal cloud embedded in
a low density transparent medium with a light source placed at one end of the computational
domain. Here we generalize that test problem by replacing the spheriodal cloud with
a density-stratified slab. As we will demonstrate, density stratification serves to exercise
the multi-frequency capabilities of our new transport algorithm. 

Four density layers are utilized in this test, along with three frequency groups.
The background (transparent medium) density is fixed at $10^{-3}$ g cm$^{-3}$,
while the opaque object is layered (from the bottom up) with densities of $10^{3}$, $10^{2}$, and
$10$ g cm$^{-3}$.
The opaque slab is 1 cm long by 1 cm tall and rests 0.5 cm along the bottom of
a grid $L=3$ cm long by $4/3$ cm tall, resolved with $384\times 192$ zones.
Each density layer is $1/3$ cm thick.
The gas and radiation begin in cold equilibrium with
$T_\mathrm{gas} = T_\mathrm{rad} = 290~\mathrm{K}$ with a gas
adiabatic index of $\Gamma = 5/3$. The photon streams are initialized
at the left boundary with a uniform (frequency-integrated) source temperature $T_{\mathrm{source}}$ = 1740 K,
so that $E = a_R T^4$ and $F^x = 0.99999 E$. Spectral energies and fluxes are
initialized as $E_{(\nu)n} = E/(h\delta\nu_n)/N_b$ and $F_{(\nu)n} = c E_{(\nu)n}$, where $N_b=3$ is the number of frequency bins.

The opacity of the gas is designed
to produce the desired behavior of the bin-center energies through
the different density layers. In particular, the frequency bin edges ($2\times10^{-3}$ and 2 eV$/h$),
the number of frequency groups (3, with logarithmic spacing), the opacity parameters, and the density layers are
tailored so that each stacked layer of the slab is optically thin to a different frequency bin so that
we can test shadow casting for each individual frequency group. This is accomplished with the
following absorption opacity power law (we do not consider scattering here):
\begin{equation} 
\kappa^{\mathrm{a}} = \kappa_0 \left(\frac{\rho}{\rho_0}\right)^{3} 
                               \left(\frac{\nu}{\nu_{0}}\right)^{-4} \mathrm{cm}^{2}~\mathrm{g}^{-1} ~,
\end{equation}
where the coefficient $\kappa_0$ is chosen to normalize the mean free path of the 
highest frequency photons through the densest (inner) layer to $\approx 10^{-2}$ cm, thus
guaranteeing that the slab is sufficiently opaque in its bottom layer to block all
streaming photons. Furthermore, the combination of densities and opacity power-law parameters
define essentially the same optical thickness to the middle frequency photons in the
middle layer, and to the lowest frequency photons in the topmost layer.

\begin{figure}
\begin{center}
\includegraphics[width=0.5\textwidth]{\figpath/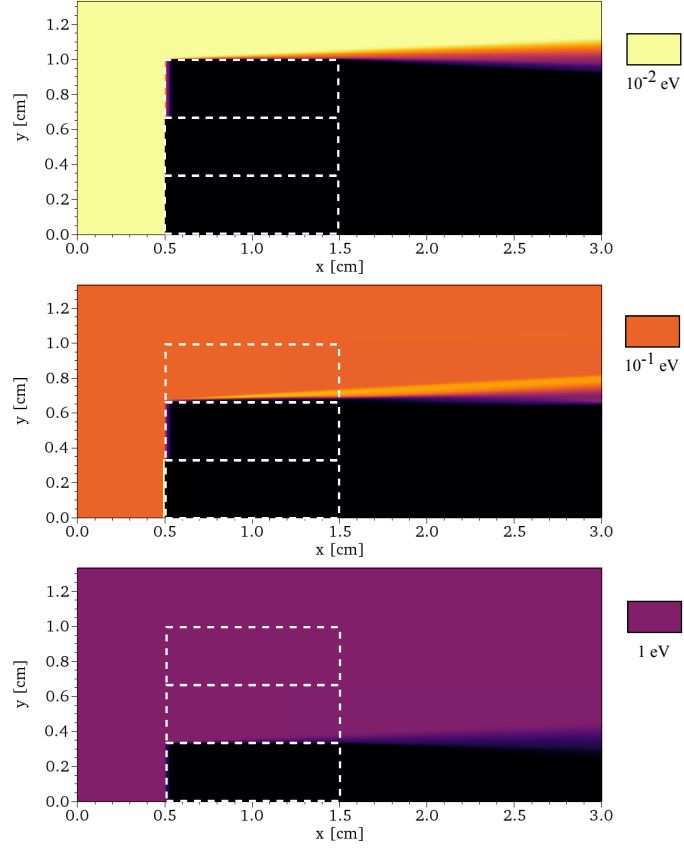}
\caption{
Pseudocolor plot of the conserved radiation spectral energy, ${\cal R}_{(\nu)n}$, for the multi-frequency shadow test.
The bin center energies, $h\nu_n$, increase from $10^{-2}$ eV (top) to
$10^{-1}$ eV (middle) to $1$ eV (bottom) images. Colors are linearly scaled, using an independent normalization for each frame. The three different slabs (dashed, white lines) are constructed to be transparent at different frequencies. 
}
\label{fig:shadow}
\end{center}
\end{figure}

\begin{figure}
\begin{center}
\includegraphics[width=0.5\textwidth]{\figpath/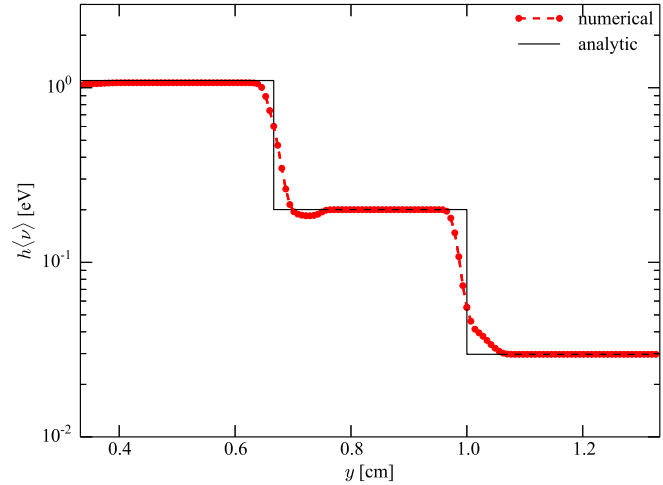}
\caption{
Line plot of the mean photon energy along the vertical ($y$) direction 
for all photons that cross position $x=1.8$ cm in the multi-frequency slab shadowing problem. 
The first slab ($y<1/3$ cm) is not shown since it is opaque to radiation.
The black line is the expected photon energy calculated by
only including source photons expected to be transmitted
at a given height. 
}
\label{fig:shadow_nu}
\end{center}
\end{figure}

From this configuration we expect to observe the following: all photons will be blocked by
the bottom (highest density) layer; photons in the two lowest frequency bins will
be blocked in the middle layer; only the lowest frequency photons will be blocked
in the upper layer; and all photons will stream through the low density
gap above the opaque slab. Hence we should see a clear separation of photon streams
and shadows if we plot each frequency bin separately.
The results, after 1.5 light-crossing times, as shown in Figure \ref{fig:shadow}, confirm these expectations.
The three images plot ${\cal R}_{(\nu)n}$ for the three photon
bin center frequencies, at roughly 0.01, 0.1 and 1 eV$/h$.
As expected, the density layers produce sharp, clear shadows in both space and
spectral energy. Notice that, as the radiation propagates, the edges of the shadows tend to flare out, a trait that is sensitive to the reconstruction method and limiter steepness \citep{davis12,mckinney14}, yet the transition from light to dark
is nevertheless quite pronounced at all frequencies. To demonstrate the clean separation of 
the three spectral components after passing through the stratified slab, we
plot in Figure \ref{fig:shadow_nu} the average photon energy as a function
of the vertical height, $y$, crossing the horizontal position $x=1.8$ cm. Also plotted is an ``analytic''
solution that is calculated by only summing the energy within the bin ranges that are theoretically transparent for a given height.
The agreement is quite good.

\subsection{Two-Beam Shadow Test}
\label{sec:2shadow}

The shadow test from the previous section demonstrates one advantage of the $\bf{M}_1$ closure scheme over simpler flux-limited diffusion, the fact that it accurately casts shadows from a single beam incident upon an opaque object. However, a well-known shortcoming of flux-integrated (grey) $\bf{M}_1$ is that intersecting beams of light do not correctly cross one another. Instead, they merge, flowing in the direction of the average, resultant flux. A traditional illustration of this is the two-beam, shadow test \citep{sadowski13,fragile14,mckinney14}. In this test, two beams of radiation enter the computational domain, one from the upper and one from the lower boundaries. Each beam is angled toward a circular cloud along the centerline of the grid. Rather than the two beams casting two independent shadows as would be expected, they cast three partial shadows, one each in the directions of the original beams, and one in the ``average'' beam direction. This third shadow is the unphysical result of the partial merging of the two beams.

\begin{figure}
\begin{center}
\includegraphics[width=0.8\textwidth]{\figpath/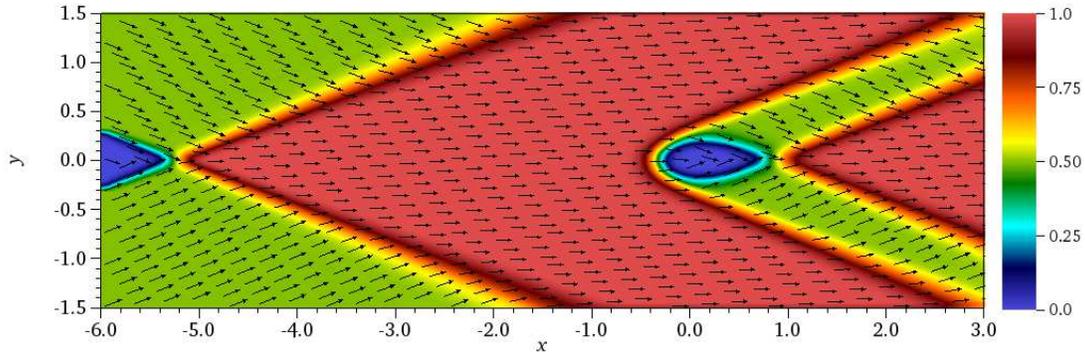}
\caption{
Pseudocolor plot of the frequency-integrated conserved radiation energy, ${\cal R}$, with vectors representing the radiation velocity, $u^i_R$, for the two-beam cloud shadow test at $t = 20$. Note the cloud casts two shadows, as it should.
}
\label{fig:2shadow}
\end{center}
\end{figure}

While our current multi-frequency $\bf{M}_1$ radiation transport does not provide a true solution to the issue of merging beams, it does admit an interesting workaround. Since photons (or beams) in different frequency bins are advected independently, and since there is no process in this test to trigger frequency exchanges, two beams of different frequencies can propagate independently, cross as expected, and cast individual shadows as they should. To illustrate this, we repeat the two-beam shadow test as presented in \citet{sadowski13,fragile14}, but with the new twist that each beam occupies a different frequency bin (the bin boundaries are not relevant). The test is run on a $120\times 120$ grid, obviously with no reflection applied at $y=0$. Figure \ref{fig:2shadow} confirms our expectation that the two beams now leave only two shadows, one in each of the beam directions. Interestingly, since Figure \ref{fig:2shadow} shows the frequency-integrated radiation energy and velocity, the beams {\em appear} to merge in the red triangular regions, but are actually still traveling in their independent directions. 

\subsection{Beam of Light Near a Black Hole}
\label{sec:beam}

An important test for a general relativistic radiation transport scheme is to verify that the radiation propagates along geodesics as expected in strong gravitational fields. To verify this, we reproduce a series of light beam tests introduced by \citet{sadowski13}.
For these tests we initialize a photon beam in the curved spacetime geometry near a
$3 M_\odot$ Schwarzschild black hole, and neglect any coupling interactions between the
gas and radiation ($\kappa^\mathrm{a} = \kappa^\mathrm{s} = 0$) so
that we can compare the path of the radiation beam against accurate geodesic paths.
In this section, all distances are measured in units of $GM/c^2$.
All calculations are run with 5 frequency groups covering 10 to $10^4$ eV$/h$ on a two-dimensional $r-\phi$ grid, with resolution $320 \times 320$
and grid coverage over $0 \le \phi \le \pi/2$ and 
$r_\mathrm{in} < r < r_\mathrm{out}$. We consider two cases:
($r_\mathrm{in}, r_\mathrm{out}, r_\mathrm{beam}$) = 
(2.5, 3.5, 3.0$\pm0.1$) and (5.5, 11.5, 6$\pm0.2$), where $r_\mathrm{beam}$ defines the
beam center and width.
Note that the beam in the first case is centered at the photon orbit radius, $r_\mathrm{beam} = r_\mathrm{p.o.} = 3$, 
meaning that photons in the center of the beam should be able to orbit the black hole indefinitely.
The radiation temperature within the initial beam is $T_\mathrm{beam} = 10 T_0 = 10^7$ K, 
where $T_0$ is the temperature of the background radiation.  
The radiation beam has an initial Lorentz factor of $\gamma = 10$ in the grid frame.
The beam initial conditions are held constant at the $\phi = 0$ boundary.

\begin{figure}
\includegraphics[width=0.5\textwidth]{\figpath/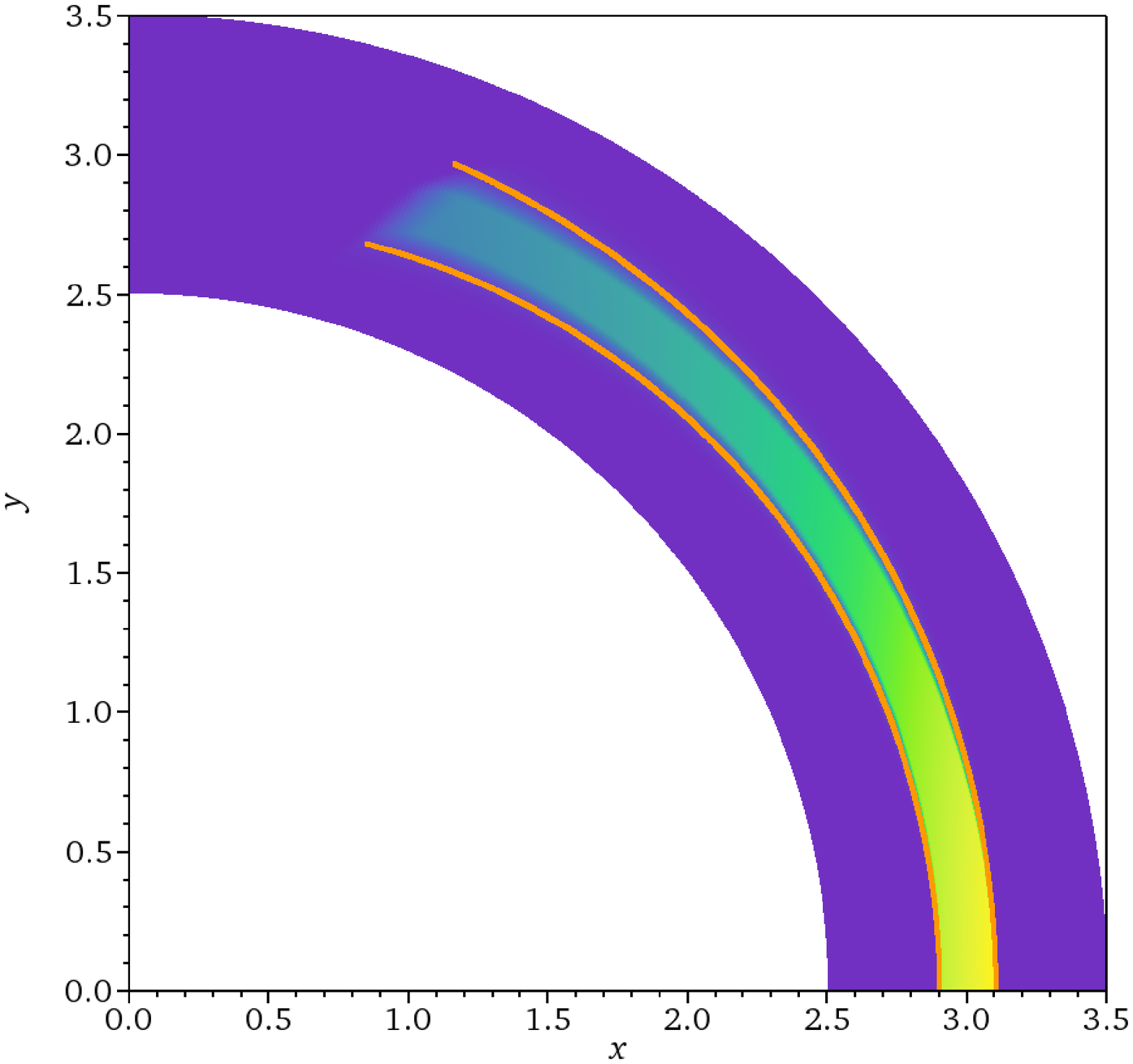}
\includegraphics[width=0.5\textwidth]{\figpath/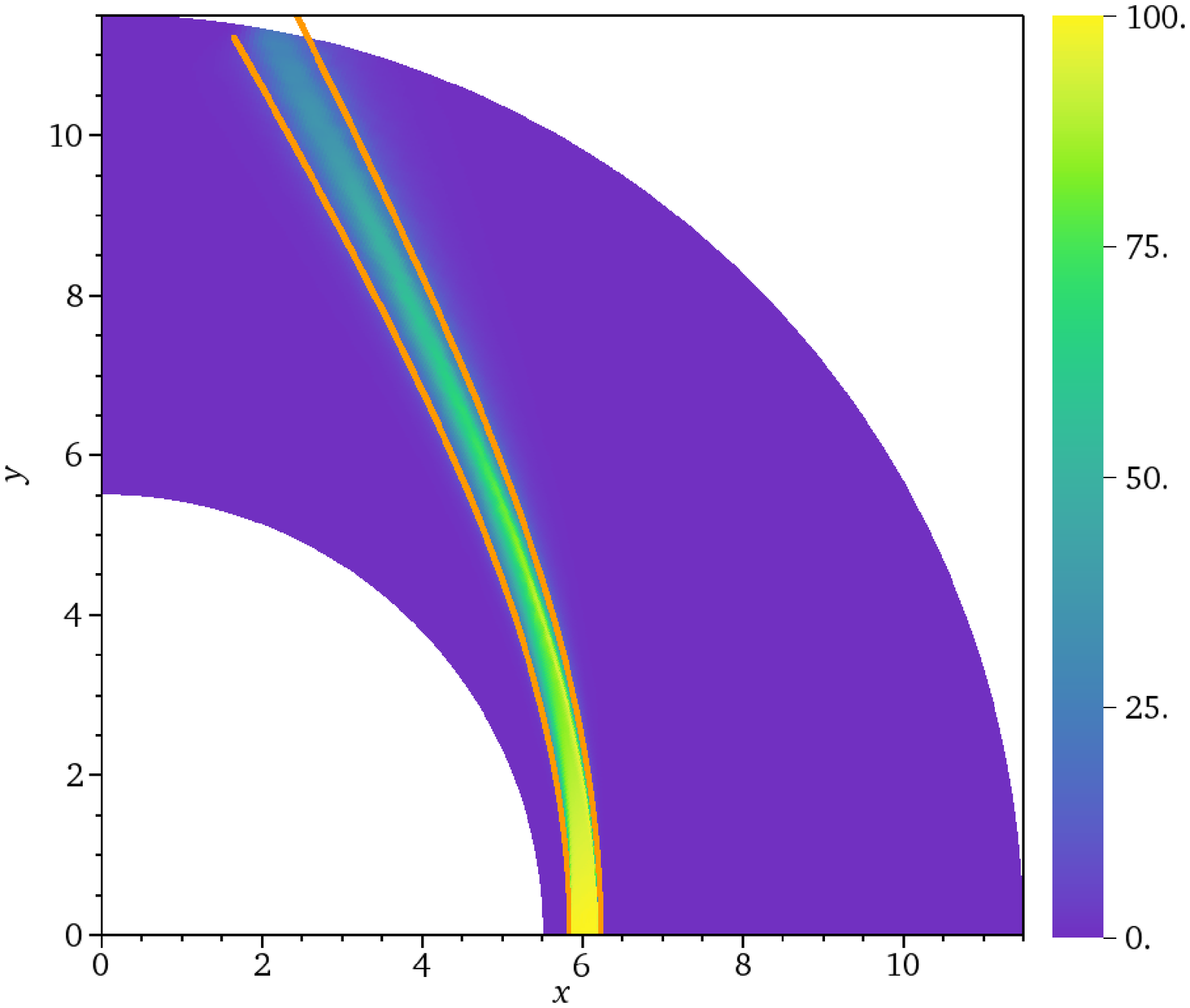}
\caption{
Pseudocolor of the frequency-integrated conserved radiation energy, $R^0_0 = \int R^0_{0(\nu)} d\nu$, (in code units) for the light beam tests.
A Schwarzschild black hole is located at the coordinate origin.  The light beams are introduced
at the bottom boundary and propagate counterclockwise around the black hole.
The orange curves represent geodesic paths starting at the initial inner and outer boundaries of the beam. Note that the background radiation energy is at least 5 orders of magnitude less than in the beam.
}
\label{fig:beam}
\end{figure}

Figure \ref{fig:beam} shows the track of each radiation beam summed over all frequency bins, i.e., $R^0_0 = \int R^0_{0(\nu)} d\nu$,
along with geodesic paths corresponding to the initial inner and 
outer boundaries of each beam. The left image corresponds to
the photon orbit case with $r_\mathrm{beam} = 3$, while the right
corresponds to $r_\mathrm{beam}=6$.
We see that each beam stays confined within the prescribed geodesic tracks and experiences the expected curvature.
Furthermore, each frequency group experiences the same curvature, so that the beam intensity is independent of the number of groups.

\section{Conclusion}
\label{sec:conclusion}

In this work, we have extended the capabilities of our {\em Cosmos++} computational astrophysics code to include multi-frequency radiation. This is done by selecting a finite number of frequency groups and independently evolving the radiation energy densities and momenta associated with each. We stick with the same explicit-implicit split of advection and radiation source terms and the two-moment $\bf{M}_1$ closure scheme as our previous work \citep{fragile14}. Thus, we retain the ability to stably evolve a large range of parameter space from optically thin to optically thick flows with reasonable time steps and accuracy, while now relaxing the grey (frequency-integrated) approximation. In this work, we have focused on presenting and testing the general relativistic version, though we note that we have also implemented a Newtonian version as well. We also have a multi-group, flux-limited-diffusion option, which we plan to report on elsewhere.

This multi-frequency capability expands the range of physical processes that can be properly captured in a simulation. For example, we demonstrated how the new code could successfully treat frequency-dependent opacities and Doppler and gravitational frequency shifts, and to some extent overcome the multi-beam shadowing limitations of the $\bf{M}_1$ closure scheme.

While multi-frequency methods are already in use in studies of core-collapse supernovae \citep[e.g.,][]{just15,kuroda16}, they have many potential uses beyond that class of problem. Possible applications include Type Ia supernovae, neutron star mergers, tidal disruption events, and super-Eddington accretion onto compact objects. Another potential application is in the study of black hole X-ray binary accretion disks, especially in the so-called intermediate spectral states, where hard and soft X-ray photons each play important roles, not only in the observed spectra, but in physically interacting with the accretion flow, affecting its structure and thermodynamic state. It is important in such an application for hard and soft X-ray photons to be able to propagate independently, experience the expected gravitational and Doppler frequency shifts, follow proper geodesic paths, and interact with the gas in a frequency-dependent manner -- all of the capabilities we have demonstrated in this paper.

\acknowledgments
The work by P.A. was performed under the auspices of the U.S. Department of Energy by Lawrence Livermore National Laboratory under Contract DE-AC52-AC52-07NA27344. P.C.F. gratefully acknowledges support from National Science Foundation grants AST-1616185 and AST-1907850. This work used the Extreme Science and Engineering Discovery Environment (XSEDE), which is supported by National Science Foundation grant number ACI-1053575.

\appendix

\section{1st Order Taylor Expansion Terms}
\label{sec:derivs}

The Jacobian matrix, ${\bf A}$, from Section \ref{sec:primitive} can either be calculated analytically or numerically.  
Although more tedious to code, we have found that the analytic method is consistently 
faster on all our tests, making it perhaps worth the extra effort.  To aid those who might 
wish to code the analytic solution, we record all the pertinent partial derivatives for 
equation (\ref{eqn:A}) here, ordered by conserved field. Notice we have dropped the frequency
subscript notation in most of these expressions, but emphasize that 
radiation related derivatives apply to all groups. 

Mass density:
\begin{align*}
  \frac{\partial D}{\partial \rho}      &= W \\
  \frac{\partial D}{\partial {u}^i}     &= \sqrt{-g} \rho \frac{\partial u^0}{\partial {u}^i} \\
  \frac{\partial D}{\partial \epsilon}  &= \frac{\partial D}{\partial E_R} = \frac{\partial D}{\partial {u}_R^i} = 0
\end{align*}

Fluid energy:
\begin{align*}
\frac{\partial {\cal E}}{\partial \rho}       &= -\sqrt{-g} \left[(1 + \epsilon) u^0 u_0 + (u^0 u_0 + 1) \frac{\partial  P_\mathrm{gas}}{\partial  \rho} \right] \\
\frac{\partial {\cal E}}{\partial  \epsilon}  &= -\sqrt{-g} \left[\rho u^0 u_0 + (u^0 u_0 + 1) \frac{\partial P_\mathrm{gas}}{\partial \epsilon} \right] \\
\frac{\partial {\cal E}}{\partial {u}^i}      &= -\sqrt{-g} (\rho + \rho \epsilon + P_\mathrm{gas}) 
                                                       \left(u_0 \frac{\partial u^0}{\partial {u}^i} + 
                                                             u^0 \frac{\partial u_0}{\partial {u}^i}\right) \\
\frac{\partial {\cal E}}{dE_R}                 &= \frac{\partial {\cal E}}{d{u}_R^i} = 0
\end{align*}

Fluid momentum:
\begin{align*}
\frac{\partial{\cal S}_j}{\partial\rho}       &= \sqrt{-g} u^0 u_j \left(1 + \epsilon + \frac{\partial P_\mathrm{gas}}{\partial \rho} \right) \\
\frac{\partial{\cal S}_j}{\partial\epsilon}   &= \sqrt{-g} u^0 u_j \left(\rho + \frac{\partial P_\mathrm{gas}}{ \partial\epsilon} \right) \\
\frac{\partial{\cal S}_j}{\partial{u}^i}      &= \sqrt{-g} (\rho + \rho \epsilon + P_\mathrm{gas}) 
                                                      \left( u_j \frac{\partial u^0}{\partial {u}^i} + 
                                                             u^0 \frac{\partial u_j}{\partial {u}^i} \right) \\
\frac{\partial{\cal S}_j}{\partial E_R}       &= \frac{\partial{\cal S}_j}{\partial{u}_R^i} = 0
 \end{align*}

Radiation energy:
\begin{align*}
\frac{\partial{\cal R}}{\partial E_R}       &= -\sqrt{-g} \left(\frac{4}{3}  u^0_R (u_R)_0 + \frac{1}{3} \right) \\
\frac{\partial{\cal R}}{d{u}^i_R}           &= -\frac{4}{3} \sqrt{-g} E_R 
                                               \left( (u_R)_0 \frac{\partial u^0_R}{\partial {u}^i_R} + 
                                                      u^0_R \frac{\partial(u_R)_0}{\partial {u}^i_R} \right) \\
\frac{\partial{\cal R}}{\partial \rho}      &= \frac{\partial{\cal R}}{\partial \epsilon} 
                                             = \frac{\partial{\cal R}}{\partial {u}^i} = 0
\end{align*}

Radiation momentum:
\begin{align*}
\frac{\partial{\cal R}_j}{\partial E_R}     &= \frac{4}{3} \sqrt{-g} u^0_R (u_R)_j  \\
\frac{\partial{\cal R}_j}{\partial{u}^i_R}  &= \frac{4}{3} \sqrt{-g} E_R 
                                                        \left( (u_R)_j \frac{\partial u^0_R}{\partial u^i_R} + 
                                                               u^0_R \frac{\partial(u_R)_j}{\partial u^i_R}\right) \\
\frac{\partial{\cal R}_j}{\partial\rho}     &= \frac{\partial{\cal R}_j}{\partial\epsilon} 
                                                      = \frac{\partial{\cal R}_j}{\partial {u}^i} = 0
\end{align*}

Also appearing in the Jacobian are the following gradients of the radiation 4-force density:
\begin{align*}
\frac{\partial G_\mu}{\partial\rho} =&
    - \left(\kappa^\mathrm{a} + \kappa^\mathrm{s}\right) R_{\mu \nu} u^{\nu} 
    - \left(\kappa^\mathrm{s} R_{\alpha \beta} u^\alpha u^\beta + \kappa^\mathrm{a} B_{(\nu)}\right) u_\mu 
    - \rho\left(R_{\mu \nu}u^\nu + B_{(\nu)} u_\mu\right) \frac{\partial\kappa^\mathrm{a}}{\partial\rho} \\
  & - \rho \left(R_{\mu \nu}u^\nu + R_{\alpha \beta} u^\alpha u^\beta u_\mu\right)\frac{\partial\kappa^\mathrm{s}}{\partial\rho} 
    - \rho \kappa^\mathrm{a} u_\mu \frac{dB_{(\nu)}}{dT} \frac{\partial T}{\partial\rho} \\
\frac{\partial G_{\mu} }{\partial\epsilon} =& 
    - \rho \left(R_{\mu \nu}u^\nu + B_{(\nu)} u_\mu\right) \frac{\partial\kappa^\mathrm{a}}{\partial\epsilon} 
    - \rho \left(R_{\mu \nu} u^\nu + R_{\alpha\beta} u^\alpha u^\beta u_\mu\right)\frac{\partial\kappa^\mathrm{s}}{\partial\epsilon} 
    - \rho \kappa^\mathrm{a} u_\mu \frac{dB_{(\nu)}}{dT} \frac{\partial T}{\partial\epsilon} \\
\frac{\partial G_\mu}{\partial{u}^i} =& 
    - \rho(\kappa^\mathrm{a} + \kappa^\mathrm{s}) R_{\mu \nu} \frac{\partial u^\nu}{\partial {u}^i} 
    - \rho \left(\kappa^\mathrm{s} R_{\alpha \beta} u^\alpha u^\beta + \kappa^\mathrm{a} B_{(\nu)} \right) \frac{\partial u_\mu}{\partial {u}^i} 
    - \rho \kappa^\mathrm{s} u_\mu R_{\alpha \beta} \left( u^\alpha \frac{\partial u^\beta}{\partial {u}^i} + u^\beta \frac{\partial u^\alpha}{\partial {u}^i}\right)\\
\frac{\partial G_\mu}{\partial E_R} =& 
    - \rho (\kappa^\mathrm{a} + \kappa^\mathrm{s}) u^\nu \frac{\partial R_{\mu\nu}}{\partial E_R} 
    - \rho \kappa^\mathrm{s} u_\mu u^\alpha u^\beta \frac{\partial R_{\alpha \beta}}{\partial E_R} \\
\frac{\partial G_\mu}{\partial {u}^i_R} =& 
    - \rho (\kappa^\mathrm{a} + \kappa^\mathrm{s}) u^\nu \frac{\partial R_{\mu\nu}}{\partial {u}^i_R} 
    - \rho \kappa^\mathrm{s} u_\mu u^\alpha u^\beta \frac{\partial R_{\alpha \beta}}{\partial {u}^i_R} \\
\end{align*}

Finally, the following partial derivatives are needed to evaluate the above expressions:
\begin{align*}
\frac{\partial u_\alpha}{\partial {u}^i}       &= g_{0\alpha} \frac{\partial u^0}{\partial {u}^i} + g_{i\alpha} \\
\frac{\partial (u_R)_\alpha}{\partial u_R^i} &= g_{0\alpha} \frac{\partial u_R^0}{\partial u_R^i} + g_{i\alpha} \\
\frac{\partial R_{\alpha \beta}}{\partial E_R}    &= \frac{4}{3} (u_R)_\alpha (u_R)_\beta + \frac{1}{3} g_{\alpha \beta} \\
\frac{\partial R_{\alpha\beta}}{\partial {u}^i_R} &= \frac{4}{3} E_R 
                                                               \left[ (u_R)_\alpha \frac{\partial (u_R)_\beta}{\partial {u}^i_R} + 
                                                                      (u_r)_\beta \frac{\partial (u_R)_\alpha}{\partial {u}^i_R}\right]
\end{align*}
\bibliographystyle{apj}

\end{document}